\documentclass{article}

\usepackage{amssymb,amsfonts,amsmath}
\usepackage{cite,enumerate,float}
\usepackage{color}
\usepackage{tikz}
\usetikzlibrary{arrows,snakes,backgrounds}
\usepackage{url}
\usepackage[vcentermath]{youngtab}

\def\be{\begin{eqnarray}}
\def\ee{\end{eqnarray}}
\def\nn{\nonumber}

\def\p{\partial}

\def\wcs{weak compositions}

\def\NS{\mathfrak{P}}

\definecolor{red}{rgb}{1,0,0}
\definecolor{orange}{rgb}{1,0.5,0}
\definecolor{violet}{rgb}{0.7,0,1}

\def\cre{\color{red}}
\def\co{\color{orange}}

\def\cg{\color{green}}
\def\cb{\color{blue}}

\hoffset= - 1.0in         

\begin{document}

\title{\vspace{1.5cm}\bf Cherednik system: eigenfunctions at generic eigenvalues
}

\author{
A. Mironov$^{b,c,d,}$\footnote{mironov@lpi.ru,mironov@itep.ru},
A. Morozov$^{a,c,d,}$\footnote{morozov@itep.ru},
A. Popolitov$^{a,c,d,}$\footnote{popolit@gmail.com}
}

\date{ }

\maketitle

\vspace{-6cm}

\begin{center}
  \hfill MIPT/TH-18/26\\
  \hfill FIAN/TD-15/26\\
  \hfill ITEP/TH-22/26\\
  \hfill IITP/TH-20/26
\end{center}

\vspace{4.5cm}

\begin{center}
$^a$ {\small {\it MIPT, Dolgoprudny, 141701, Russia}}\\
$^b$ {\small {\it Lebedev Physics Institute, Moscow 119991, Russia}}\\
$^c$ {\small {\it NRC ``Kurchatov Institute", 123182, Moscow, Russia}}\\
$^d$ {\small {\it Institute for Information Transmission Problems, Moscow 127994, Russia}}
\end{center}

\vspace{.1cm}

\begin{abstract}
Symmetric Macdonald polynomials of $N$ variables provide eigenfunctions of the $N$-body trigonometric Ruijsenaars-Schneider integrable system at particular eigenvalues. In order to construct eigenfunctions with arbitrary eigenvalues, M. Noumi and J. Shiraishi used a recursion in $N$ (branching rule) for the symmetric Macdonald polynomials and analytically continued them. This generated a power series, which is a part of triad (universal solution). In the present paper, we demonstrate that a similar procedure is available for another integrable system, $N$-body Cherednik integrable system inspired by the DAHA of type $A$, which has non-symmetric Macdonald polynomials as its polynomial eigenfunctions. However, in this system, the generic eigenfunction is more complicated: it is not just a simple power series as in the Noumi-Shiraishi case, but has an involved structure with $N!$ branches, each of them being a sum over the Weyl chambers of power series of the Noumi-Shiraishi type. As an illustration, we also provide explicit formulas for particular cases.
\end{abstract}

\bigskip

\newcommand\smallpar[1]{
  \noindent $\bullet$ \textbf{#1}
}

\section{Introduction}

In this paper, we continue our study of integrable systems associated with Ding-Iohara-Miki (DIM) algebra \cite{DI,M} and DAHA \cite{Che}. These systems contain the trigonometric Ruijsenaars-Schneider \cite{RS} and Cherednik \cite{Chebook} systems and their generalizations. Our main point of interest here is eigenfunctions of the commuting Hamiltonians of these systems.

The general context regarding these eigenfunctions is as follows \cite{NSM1,NSM2,NSM3,NSM4}.

\begin{itemize}

\item{}
The DIM algebra or, equivalently, the elliptic Hall algebra \cite{K,BS,S} (which is basically the same \cite{S,Feigin}) possesses a family of commutative subalgebras, labeled by rational parameter $a$
and represented by ``rays'', which are connected by rotation operators
$\hat{\cal O}_{v,h}$ \cite{MMP}.
Each ray provides a family of commuting Hamiltonians.

\item{} Representation theory of the DIM algebra for a particular $N$-body representation and integer $a$ (we call the corresponding rays integer)
expresses these Hamiltonians as $a$-twisted Cherednik operators,
which act by a combination of finite difference operators in $x$ \cite{MMP}. This realization is due to equivalence of the DIM algebra and the spherical DAHA \cite{dFK1} (i.e. DAHA acting on the space of symmetric functions).

\item{} Hence, one naturally associates with any integer ray commuting Hamiltonians from DAHA, which we call $a$-twisted Cherednik Hamiltonians.

\item{} The common eigenfunctions of the DIM Hamiltonians at generic $a$ are \cite{CF,MMP1} $a$-twisted Baker-Akhiezer (BA) functions \cite{ChE}, which can be constructed only at $q=t^{-m}$, $m\in\mathbb{Z}_{\ge 0}$, and are polynomials in $x$.

\item{}
The common eigenfunctions of the twisted Cherednik (i.e. DAHA) Hamiltonians at generic $a$ can be also constructed at $q=t^{-m}$, $m\in\mathbb{Z}_{\ge 0}$, $E_\lambda^{(a,m)}$, and are polynomials in $x$, which can be represented as
$\Xi$-expansions with non-factorized rational coefficients $F_\lambda^\mu(\vec x)$
\be
E^{(a,m)}_\lambda(\vec x) = \sum_{\mu\le\lambda} F_\lambda^\mu(\vec x) \cdot \Xi^{(a,m)}_\mu(\vec{x)}
\label{Xideco}
\ee
where $\lambda$, $\mu$ are weak compositions (sets of unordered non-negative integers) of length $N$, and the Bruhat order is implied (see details in \cite{NSM4}).
Functions $\Xi_\mu$ are all made from a single but still poorly understood ``vacuum state''
$\Omega^{(a,m)}(\vec X)$, which is unclear how to lift to an arbitrary $t$. In variance with
$\Omega^{(a,m)}(\vec X)$, the coefficients $F$ are independent of $a$, and can be raised to functions of $t$,
in this sense they are rational functions of $q^{-m}$.
They can be decomposed in comprehensible and rather simple combinations of completely factorized rational functions of $x_i$.

\item{}
At $a=1$, the vacuum state $\Omega=1$ and (\ref{Xideco}) simplifies drastically.
In particular, one can immediately analytically continue formulas in $m$, and obtain a family of $t$-dependent
polynomials $E_\lambda(\vec x|q,t):=E^{(1)}_\lambda$, which are nothing but the non-symmetric Macdonald polynomials
introduced already in \cite{Opd95,Mac96,Che95}. This will be the main point of our interest here so that we will not deal with a generic $\Omega^{(a,m)}(\vec X)$ in the present text.

\item{} Similarly, the DIM Hamiltonians of the $a=1$ ray become simply rotated Ruijsenaars-Schneider Hamiltonians, and their eigenfunctions at arbitrary $t$ and peculiar eigenvalues are symmetric Macdonald polynomials $M_\lambda$, $\lambda$ being the Young diagram, or, at arbitrary eigenvalues but peculiar $t=q^{-m}$, untwisted BA (quasi)polynomial functions \cite{Cha}. Both these systems of polynomials can be obtained by reductions from the Noumi-Shiraishi (NS) power series \cite{NS} giving the eigenfunction with arbitrary eigenvalue and $t$.
    The NS function along with these two polynomial reductions form a triad \cite{MMP3}.

\item{} One can add to the symmetric Macdonald polynomials $M_\lambda$ the skew symmetric Macdonald polynomials $M_{\lambda/\mu}$ in accordance with
\be
M_R(\vec x,\vec y)=\sum_\mu M_{R/Q}(\vec y)M_Q(\vec x)
\ee
One can consider the case of only one of $y_i$ non-zero
\be
M_R(x_1,\ldots,x_{n-1},y)=\sum_Q M_{R/Q}(y)M_Q(x_1,\ldots,x_{n-1})
\ee
where the sum runs over Young diagrams $Q$ with no more than $N-1$ lines.

\item{}
In the case of non-symmetric Macdonald polynomials, one could again introduce the skew non-symmetric polynomials $E_{\lambda/\mu}$ via the expansion
\be\label{skewE}
E_\lambda(\vec x,\vec y)=\sum_\mu E_{\lambda/\mu}(\vec y)E_\mu(\vec x)
\ee
One can similarly study the skew polynomials of one variable in this case, though, in this case, the answer for $E_{\lambda/\mu}(\vec y)$ depends on which of the variables is chosen distinguished.

\item{} The $x$-independent {\it coefficients} of skew non-symmetric Macdonald polynomials of one variable $E_{\lambda/\mu}(y)$ are made
from the coefficients $F$, which are rational in $x$, but inverse operation is unavailable.
Still these coefficients are of their own value and interest, and allow one to construct a recursion procedure.

\item{}
The central {\bf new result of this paper} is that  $E_{\lambda/\mu}(q,t)$ turn out to be factorized (with no {it a priori} reason to), and we provide an explicit formula for them in terms of the $q$-Pochhammer symbols.
This allows one to analytically continue these functions to arbitrary complex values of $\lambda$.

\item{}
Successively repeating   expansion (\ref{skewE}) for
$x_N\to x_{N-1}\to\ldots\to x_1$,
we obtain a generic formula for
\be\label{Ex}
E_{\lambda}(\vec x) = \sum_\mu {\cal E}_{\lambda/\mu} x^{\mu}
\ee
where $x^{\mu}:=\prod_{i=1}^N x_i^{\mu_i}$, and $x$-independent coefficients ${\cal E}_{\lambda/\mu}$ which are $N$-fold combinations
of factorized quantities $E_{\lambda/\mu}$:
\be
{\cal E}_{\lambda/\mu} =
\sum_{\mu_1,\ldots,\mu_N} E_{\lambda/\mu_1}E_{\mu_1/\mu_2}\ldots E_{\mu_{n-1}/\mu}
\ee

\item{}
This provides a very efficient way to calculate non-symmetric Macdonald polynomials.

\item{}
Since these expressions are explicit analytic formulas (finite sums of products),
they can be analytically continued to arbitrary complex values of $\lambda$
(with summation still over \wcs $\mu_1,\ldots,\mu_{n-1},\mu$).
The resulting expression provides a generalization of the Noumi-Shiraishi
power series to the non-symmetric case.
Note that, at non-integer $\lambda$, the sum in (\ref{Ex}) becomes infinite, and the polynomial converts to power series. A sum of such power series provides eigenfunctions of the Cherednik Hamiltonians with arbitrary eigenvalues. Our goal here is to construct and briefly discuss these power series and the corresponding eigenfunctions, which can be called following \cite{dFK2} {\bf universal solutions}.

\item{} Similarly to the NS case, the universal solution in the non-symmetric case has two polynomial reductions: to non-symmetric Macdonald polynomials and to (quasi)polynomial counterparts of the BA functions. The universal solution with these two polynomial reductions forms {\bf a non-symmetric triad}.

\end{itemize}

The paper is organized as follows. In section 2, we introduce the notion of skew non-symmetric polynomials and illustrate it with a few simple examples that demonstrate features of such polynomials essential for future consideration. In section 3, we describe the integrable Hamiltonians whose eigenfunction throughout the paper we consider: these are the Ruijsenaars-Schneider Hamiltonians (their eigenfunctions, the symmetric Macdonald polynomials are also considered in this section) and Cherednik Hamiltonians. In section 4, we discuss the skew non-symmetric Macdonald polynomials at $N=2$ and $N=3$, and, in section 5, at generic $N$. At last, in section 6, we discuss the non-symmetric triad. Section 7 contains some concluding remarks.

\paragraph{Notation.} Throughout the text, we use notation Greek letters $\lambda$, $\mu$, etc. to denote \wcs and use capital Roman letters $R$, $Q$, etc to denote Young diagrams, i.e. \wcs put in a weakly descending order.

We define the $q$-Pochhammer symbol
\be
(x;q)_n:=\prod_{j=0}^{n-1}(1-xq^j)
\ee
the quantum numbers
\be
[n]_q:={q^n-1\over q-1}
\ee
and $\displaystyle{\binom{n}{k}_q}$ denotes the corresponding $q$-binomial coefficient defined to vanish when $n<k$.

\section{Skew non-symmetric polynomials: preliminary considerations}

\subsection{Generalities}

Let us consider a subset $V$ of a set $U=[1,\ldots,N]$, consisting of $m$ elements. Then, any symmetric polynomial of $N$ variables, for instance, the Schur polynomial $S_R$ can be expanded into the basis of symmetric polynomials of $\{x_i\}$, $i\in V$:
\be
S_R[\vec y \cup \vec x] = \sum_Q S^{(U/V)}_{R/Q}[\vec y]S_Q^{(V)}[\vec x]
\ee
The polynomials $S^{(U/V)}_{R/Q}[\vec y]$ are called skew (Schur) polynomials, and they depend only on two Young diagrams $R$ and $Q$, and on $m$, thanks to the symmetricity, do not depend on $U/V$.

This definition is also applicable to non-symmetric polynomials, however, in this case, the skew polynomials also depend on the choice of $V$. Hence, though one denotes the non-symmetric Macdonald polynomials just as $E_\lambda$, we denote their skew version as $E_{\lambda/\mu}^{(i_1,\ldots,i_m)}$, where $U/V=[i_1,\ldots,i_m]$.

In this paper, we introduce and study these new skew non-symmetric Macdonald  polynomials.
We begin with the case of one-dimensional case of $U/V$, i.e. with $E^{(i)}_{\lambda/\mu}$, and illustrate it with different choices of $i$: we provide just two simple examples, how one particular non-symmetric polynomial is decomposed with $N$ different choices of $U/V$.

In fact, in this paper, we mostly deal with
this particular one-dimensional case in order to construct a recursive procedure of building non-symmetric Macdonald polynomials in the number of variables $N$.

\subsection{Simple examples}

We start with the $N=2$ example:
\be
E_{[20]}(x_1,x_2) = \frac{q^2(t-1)}{q^2t-1} x_2^2 + \frac{q(t-1)(q+1)}{q^2t-1} x_1x_2 +x_1^2=
\left\{\begin{array}{c}
\frac{q^2(t-1)}{q^2t-1} x_2^2 + \frac{q(t-1)(q+1)}{q^2t-1} x_2E_{[1]}(x_1)+E_{[2]}(x_1),\
\\
\\
x_1^2+\frac{q(t-1)(q+1)}{q^2t-1}x_1 E_{[1]}(x_2) + \frac{q^2(t-1)}{q^2t-1} E_{[2]}(x_2)
\end{array}
\right.
\ee
where $E_{[r]}(x) = x^r$. This manifests the relation between expansion coefficients
\be
E^{(1)}_{\lambda/\mu} = E^{(2)}_{\lambda/\nu} \ \ \ \  {\rm for} \ \ \  \mu+\nu=|\lambda|
\ee
Likewise
\be
E_{[10]}(x_1,x_2) = \frac{q(t-1)}{qt-1}x_2+x_1, \ \ \ \ E_{[01]}(x_1,x_2) = x_2, \ \ \ \ \
E_{[02]}(x_1,x_2) = x_2^2 +\frac{t-1}{qt-1}x_1x_2
\ee
and
\be
E_{[30]}(x_1,x_2) = \frac{q^3(t-1)}{q^3t-1}x_2^3 + \frac{q^2(q^2+q+1)(t-1)(qt-1)}{(q^2t-1)(q^3t-1)}x_1x_2^2
+ \frac{q(q^2+q+1)(t-1)}{q^3t-1}x_1^2x_2  +x_1^3, \nn \\
E_{[03]}(x_1,x_2) = x_2^3 + \frac{(t-1)(q+1)}{q^2t-1}x_1x_2^2 + \frac{t-1}{q^2t-1}x_1^2x_2
\ee
Clearly all expansions exist, moreover, one can observe that $1$-dimensional skew polynomials
are {\bf nicely factorized functions of $q$ and $t$}.
This is true for expansions of arbitrary $E_\lambda$ at all levels and all $N$,
and this is going to be our {\bf first} observation.

\bigskip

{\bf Second}, we provide a generic formula for these factorized 1-dimensional polynomials
for arbitrary $\lambda$.

\bigskip

{\bf Third}, one can now iterate this expansion and get a representation of
$k$-dimensional skew polynomials as a $k$-linear combination of {\it factorized} $1$-dimensional ones.
For $k>1$, they are no longer factorized, still it is a very simple and straightforward formula.
For $k=N$, it provides decomposition(s) of original non-symmetric Macdonald polynomial,
which actually appears to be a {\bf fastest way to calculate them}.

For illustration, at $N=3$,
there are three different expansions

{\footnotesize
\be
E_{[030]}(x_1,x_2,x_3) &=& \frac{q^3(t-1)}{q^3t^2-1}x_3^3
+ \frac{q^3t(t-1)^2(q+1)}{(q^2t-1)(q^3t^2-1)}x_3^2E_{[10]}(x_1,x_2)
+ \frac{q^2(t-1)(qt^2-1)(q^3t-1)}{(qt-1)(q^2t-1)(q^3t^2-1)}x_3^2E_{[01]}(x_1,x_2) + \nn \\
&& + \frac{q^3t(1-t)^2}{(q^2t-1)(q^3t^2-1)}x_3E_{[20]}(x_1,x_2)
+ \frac{q(t-1)(q+1)(q^2t^2-1)(q^3t-1)}{(q^2t-1)^2(q^3t^2-1)} x_3E_{[02]}(x_1,x_2)
+ E_{[03]}(x_1,x_2),
\nn \\ \nn \\ \nn \\
E_{[030]}(x_1,x_2,x_3) &=&  x_2^3
+ \frac{ (t-1) (q+1)}{q^2t-1 }x_2^2E_{[10]}(x_1,x_3)
+ \frac{qt(t-1)(q^3t-1)(q^3t-1)}{(qt-1)(q^2t^3-1)}x_2^2E_{[01]}(x_1,x_3) + \nn \\
&& + \frac{(t-1)}{(q^2t-1)}x_2E_{[20]}(x_1,x_3)
+ \frac{q^2t(t-1)(q+1)(qt-1)(q^3t-1)}{(q^2t-1)^2(q^3t^2-1)} x_2E_{[02]}(x_1,x_3)
+ \frac{q^3t(t-1)}{q^3t^2-1}E_{[03]}(x_1,x_3),
\nn \\ \nn \\  \nn \\
E_{[030]}(x_1,x_2,x_3) &=&
0\cdot x_1^3 +
\frac{ (t-1) }{q^2t-1 }x_1^2E_{[10]}(x_2,x_3)
- \frac{q(t-1)^2}{(qt-1)(q^3t^2-1)}x_1^2E_{[01]}(x_2,x_3)
+ \frac{(t-1)(q+1)}{(q^2t-1)}x_1E_{[20]}(x_2,x_3) - \nn \\
&& - \frac{q^2(t-1)^2(q+1)(qt-1) }{(q^2t-1)^2(q^3t^2-1)} x_1E_{[02]}(x_2,x_3)
-\frac{q^3(t-1)^2}{(q^3t-1)(q^3t^2-1)} E_{[03]}(x_2,x_3) + E_{[30]}(x_2,x_3)
\ \ \ \ \ \
\nn
\ee
}

\noindent
of one and the same

{\footnotesize
\be
E_{[030]}(x_1,x_2,x_3) = 0\cdot x_1^3+ \frac{t-1}{q^2t-1}x_1^2x_2 + \frac{q^3t(t-1)^2}{(q^2t-1)(q^3t^2-1)}x_1^2x_3+\nn \\
+\frac{(t-1)(q+1)}{q^2t-1}x_1x_2^2+\frac{q(t-1)^2(q+1)(q^2t+qt+1)}{(q^2t-1)(q^3t^2+1)}x_1x_2x_3
+ \frac{q^3(q+1)(t-1)^2}{(q^2t-1)(q^3t^2-1)}x_1x_3^2 + \nn \\
+ x_2^3+\frac{q(t-1)(q^4t^2+q^3t^2+q^2t^2-q^2t-q-1)}{(q^2t-1)(q^3t^2-1)}x_2^2x_3
+ \frac{q^2(t-1)(q^3t^2+q^2t^2-q^2t+qt^2-qt+1)}{(q^2t-1)(q^3t^2-1)}x_2x_3^2
+ \frac{q^3t(t-1)}{q^3t^2-1}x_3^3
\ee
}

\noindent
We see that there are numerous {\it un}factorized coefficients,
which can be decomposed into sums of factorized ones in at least three different ways
dictated by the above decompositions.
For example, the three decompositions of the coefficient in front of $x_2x_3^2$ are:
{\footnotesize
\be
 \frac{q^2(t-1)(q^3t^2+q^2t^2-q^2t+qt^2-qt+1)}{(q^2t-1)(q^3t^2-1)}x_2x_3^2 = \nn \\
= \frac{q^3t(t-1)^2(q+1)}{(q^2t-1)(q^3t^2-1)}x_3^2\cdot \frac{q(t-1)}{qt-1}x_2
+ \frac{q^2(t-1)(qt^2-1)(q^3t-1)}{(qt-1)(q^2t-1)(q^3t^2-1)}x_3^2\cdot x_2 = \nn \\
=  \frac{(t-1)}{(q^2t-1)}x_2\cdot \frac{q^2(t-1)}{q^2t-1} x_3^2
+ \frac{q^2t(t-1)(q+1)(qt-1)(q^3t-1)}{(q^2t-1)^2(q^3t^2-1)} x_2\cdot x_3^2 = \nn \\
= -\frac{q^3(t-1)^2}{(q^3t-1)(q^3t^2-1)} \cdot \frac{(t-1)(q+1)}{q^2t-1}x_2x_3^2 +
\frac{q^2(q^2+q+1)(t-1)(qt-1)}{(q^2t-1)(q^3t-1)}x_2x_3^2
\ee
}

\noindent
Now the underlying relations, which guarantees the equivalence of three expansions, are more involved:
\be
\sum_{\mu,\nu} E^{(i)}_{\lambda/\mu}E^{(j)}_{\mu/\nu} x_i^{|\lambda|-|\mu|}x_j^{|\mu|-|\nu|}x_k^{|\nu|}
\ \ \ \text{do not depend on the choice of } i, j
\ee
i.e. is a kind of associativity relation. Clearly, in general, there are $N!$ different decompositions.

\bigskip

{\bf Forth}, this provides a unified formula for all $E_\lambda$ at a given $N$.

However, this unification is not full:
actually the formulas are different for different orderings of $\lambda_i$'s.
E.g. at $N=2$ (see details in sec.\ref{secN2}), there is one formula for $\lambda_1\geq \lambda_2$ and another one for $\lambda_2>\lambda_1$.
At $N>3$, the common interpolation exists for all $E_{[\lambda_1,\lambda_2,\lambda_3]}$
with, say, $\lambda_2>\lambda_1\geq \lambda_3$,
but formulas for $E_{[300]}$, $E_{[030]}$ and $E_{[003]}$ are different.
In general, there are $N!$ formulas for each $N$. Since each function vanishes outside its proper domain, they can be treated either as branches of one unified function, or as an $N!$-component vector of solutions to the eigenfunction equations for the Cherednik Hamiltonians.

\bigskip

{\bf Fifth}, these unified formulas are expressed through $q$-Pochhammer factors and thus
can be easily analytically continued from integer to arbitrary complex values of $\lambda$.
However, the sums over integer-valued $\mu$ are no longer cut-off at $\mu=\lambda$,
thus they become infinite, thus giving rise to {\it series} instead of polynomials.
These series are a generalization of the Noumi-Shiraishi series \cite{NS} to non-symmetric case,
i.e. we obtain a generalization of Cherednik eigenfunctions to series with
complex eigenvalues.

\section{Constructing the recursion procedure in $N$}

\subsection{Ruijsenaars-Schneider integrable system and symmetric Macdonald polynomials}

In this section, we are going to construct a recursion procedure in the number of $x_i$.
First of all, let us consider how it works in the symmetric case. We basically follow \cite{NS}.

The symmetric Macdonald polynomials are eigenfunctions of the Ruijsenaars-Schneider Hamiltonians \cite{RS}
\be\label{RH}
H_k^{RS}=\sum_{i_1<i_2<\ldots<i_k}{\left[\prod_{m=1}^kt^{\hat D_{i_m}}\Delta(x)\right]\over\Delta(x)}
\prod_{m=1}^k q^{\hat D_{i_m}}
\ee
where $\Delta(x)=\prod_{i<j}(x_i-x_j)$ is the Vandermonde determinant, and $\hat D_i:=x_i{\p\over\p x_i}$. Note that $H^{RS}_k=0$ at $k>n$ automatically. The first of the Hamiltonians is the celebrated Macdonald-Ruijsenaars operator \cite{MRS}
\be\label{MR}
H^{MR}=\sum_{i=1}^N\prod_{j\ne i}{tx_i-x_j\over x_i-x_j}q^{\hat D_i}
\ee
The symmetric Macdonald polynomials $M_R(x)$ are defined as polynomial eigenfunctions of $H^{MR}$ with eigenvalues
\be
\Lambda_R=\sum_iq^{R_i}t^{N-i}
\ee
They are enumerated by arbitrary Young diagrams $R$.

One can add to the symmetric Macdonald polynomials $M_R$ the skew symmetric Macdonald polynomials $M_{R/Q}$ in accordance with
\be\label{Msk}
M_R(\vec x,\vec y)=\sum_Q M_{R/Q}(\vec y)M_Q(\vec x)
\ee
where there are $N-1$ variables $x_i$ and $m$ variables $y_i$ so that the sum runs over Young diagrams $Q$ with no more than $N-1$ lines.

Now consider the case of only one non-zero $y$, i.e. $m=1$. Then, the branching rule is \cite{Macbook,NS}
\be\label{Mnn-1}
M_{R/Q}(y)=y^{|R|-|Q|}\times\prod_{i<j}{(q^{Q_i-R_j+1}t^{j-i-1};q)_{R_i-Q_i}\over (q^{Q_i-R_j}t^{j-i};q)_{R_i-Q_i}} \times\prod_{i\le j\le n-1}{(q^{Q_i-Q_j}t^{j-i+1};q)_{R_i-Q_i}\over(q^{Q_i-Q_j+1}t^{j-i};q)_{R_i-Q_i}}
\ee
If one associates this $y$ with the $N$-th symmetric variable, one gets an expression for the Macdonald polynomial of $N$ variables
\be\label{recM}
M_R(x_1,\ldots,x_{N-1},y)=\sum_{Q:\ l_Q\le N-1} M_{R/Q}(y)M_Q(x_1,\ldots,x_{N-1})=
\sum_{Q:\ l_Q\le N-1} y^{|R|-|Q|}M_{R/Q}(1)M_Q(x_1,\ldots,x_{N-1})
\ee
and the sum runs over Young diagrams $Q$ with no more than $N-1$ lines.
Thus, one obtains a convenient recursion (\ref{recM}) with factorized coefficients (\ref{Mnn-1}) in the number $N$ of variables in Macdonald polynomials.

Now one can use this recursion in order to start with the trivial case of $N=1$ and generate polynomials at any $N$. Moreover, one can analytically continue formula (\ref{Mnn-1}) to arbitrary complex values of $R_i$'s, and generate this way eigenfunctions of the Macdonald-Ruijsenaars operator with arbitrary eigenvalues. However, in this case, the sum in (\ref{recM}) is no longer finite, and one has a power series \cite{NS} instead of the Macdonald polynomial.

\subsection{Cherednik integrable system and non-symmetric Macdonald polynomials\label{3.2}}

Now we consider the Cherednik integrable system, its commuting Hamiltonians $C_i$, $i=1,\ldots,N$ are defined
\be\label{Cpi}
C_i=t^{N-i}T_iT_{i+1}\ldots T_{N-1}\pi T_1^{-1}T_2^{-1}\ldots T_{i-1}^{-1}
\ee
with
\be\label{TxA}
T_i=1+{x_i-t^{-1}x_{i+1}\over x_i-x_{i+1}}(\sigma_{i,i+1}-1),\ \ \ \ \ i=1,\ldots,N-1
\ee
where $\sigma_{i,j}$ permutes $x_i$ and $x_j$, and
\be\label{pix}
\pi F(x_1,x_2,\ldots,x_N)=F(qx_N,x_1,\ldots,x_{N-1})
\ee
Note that the Ruijsenaars-Schneider Hamiltonians, which are elements of a commutative subalgebra of the DIM algebra in the $N$-body representation \cite{MMP} can be constructed from the Cherednik Hamiltonians in the $x$-representation restricting the space of functions they are acting on:
\be
H_k^{RS}=\sum_{i=1}^NC_i^k\Big|_{symm}
\ee
where $\Big|_{symm}$ means the space of symmetric functions of $N$ variables $\{x_i\}$.
This reflects a correspondence between the DIM algebra and spherical DAHA \cite{dFK1}.

Since the Cherednik Hamiltonians are not symmetric in variables $x_i$, their common polynomial eigenfunctions are expectedly non-symmetric. That is, they are celebrated non-symmetric Macdonald polynomials defined as polynomial eigenfunctions of $C_i$'s with eigenvalues
\be\label{evw}
\Lambda^{(i)}_{\lambda}&=&q^{\lambda_i}t^{N-1-\zeta_\lambda(i)}\nn\\
\zeta_\lambda(i):&=&\#\{k<i|\lambda_k\ge\lambda_i\}+\#\{k>i|\lambda_k>\lambda_i\}
\ee
They are enumerated by arbitrary weak compositions $\lambda$.

In the case of non-symmetric Macdonald polynomials, one could again introduce the skew non-symmetric polynomials $E_{\lambda/\mu}$ via the expansion
\be\label{skewns}
E_\lambda(\vec x,\vec y)=\sum_\mu E_{\lambda/\mu}(\vec y)E_\mu(\vec x)
\ee
However, this expansion and the skew polynomial $E_{\lambda/\mu}(\vec y)$ in this case depend on the relative positions of $x_i$'s and $y_j$'s.

In particular, one can repeat what has been done in the symmetric case and choose one variable $y$ and $N-1$ variables $x_i$. This would give
\be\label{recE}
E_\lambda(x_1,\ldots,x_{k-1},y,x_{k+1},\ldots,x_{N})&=&\sum_{\mu:\ l_\mu\le N} E_{\lambda/\mu}^{(k)}(y)E_\mu(x_1,\ldots,x_{k-1},x_{k+1},\ldots,x_N)
\ee
The sum here runs over the weak compositions $\mu$ with $N-1$ elements, giving rise to the recursion $N$, and
\be
E_{\lambda/\mu}^{(k)}(y)=y^{|\lambda|-|\mu|}E_{\lambda/\mu}^{(k)}(1)
\ee
The skew coefficients $E_{\lambda/\mu}^{(k)}(1)$'s are factorized!

Note that $E_{\lambda/\mu}^{(k)}(1)$ at different $k$ are related with each other. To understand this, we use that
the non-symmetric Macdonald polynomials enjoy the property (Knop–Sahi recurrence) \cite{KS,HHL}
\be\label{B}
E_{[\lambda_2,\ldots,\lambda_N,\lambda_1+1]}(x_1,x_2,\ldots,x_N)=
q^{-\lambda_1}x_NE_{[\lambda_1\lambda_2,\ldots,\lambda_N]}(qx_{n},x_1,x_2,\ldots,x_{N-1})
\ee
From this formula and (\ref{recE}), it follows that
\be
E_{[\lambda_1,\ldots,\lambda_N]}(x_1,x_2,\ldots,x_N)&=&q^{\lambda_1+1}x_1^{-1}
E_{[\lambda_2,\ldots,\lambda_N,\lambda_1+1]}(x_2,\ldots,x_N,q^{-1}x_1)=\nn\\
&=&q^{\lambda_1+1}x_1^{-1}\sum_{\mu:\ l_\mu\le N-1} E_{[\lambda_2,\ldots,\lambda_N,\lambda_1+1]/\mu}^{(N)}(q^{-1}x_1)E_{\mu}(x_2,\ldots,x_N)=\nn\\
&=&q^{|\mu|-|\lambda|+\lambda_1}
\sum_{\mu:\ l_\mu\le N-1}x_1^{|\lambda|-|\mu|} E_{[\lambda_2,\ldots,\lambda_N,\lambda_1+1]/\mu}^{(N)}(1)E_{\mu}(x_2,\ldots,x_N)
\ee
i.e.
\be\label{x1n}
E^{(1)}_{\lambda/\mu}(1)=q^{|\mu|-|\lambda|+\lambda_1}E_{[\lambda_2,\ldots,\lambda_N,\lambda_1+1]/\mu}^{(N)}(1)
\ee
One can similarly obtain all other $E^{(k)}_{\lambda/\mu}(1)$ using other moves from the Knop-Sahi recurrence \cite{KS,HHL}. From now on, for the sake of definiteness, we make a recursion in $x_N$, keeping in mind that one can immediately obtain from it $E^{(k)}_{\lambda/\mu}$ for any $k$. Hence, from now on, we omit the superscript $(N)$ from the skew polynomials $E$.

\section{Skew non-symmetric Macdonald polynomials at particular $N$}

\subsection{Skew non-symmetric Macdonald polynomials at $N=2$\label{secN2}}

Though our goal is to construct a universal solution for all $E_\lambda$ at a given $N$, it is not given by a unique formula: actually the formulas are different for different orderings of $\lambda_i$'s.

For example, at $N=2$, there are two cases
\be\label{E21i}
E^>_{\lambda_1,\lambda_2} = (x_1x_2)^{\lambda_2}  \sum_{k=0}^{\lambda_1-\lambda_2} x_1^kx_2^{\lambda_1-\lambda_2-k}
q^{\lambda_1-\lambda_2-k}
\binom{\lambda_1-\lambda_2}{k}_q
\prod_{i=0}^k \frac{1-q^it}{1-q^{\lambda_1-\lambda_2-i}t}
\ \ \ \ {\rm for} \ \ \lambda_1\geq \lambda_2
\ee
and
\be\label{E12i}
E^<_{\lambda_1,\lambda_2}   = (x_1x_2)^{\lambda_1} \sum_{k=0}^{\lambda_2-\lambda_1-1} x_1^kx_2^{\lambda_2-\lambda_1-k}
\binom{\lambda_2-\lambda_1-1}{k}_q
\prod_{i=0}^k \frac{1-q^{i-1}t}{1-q^{\lambda_2-\lambda_1-i}t}
\ \
\ \ {\rm for} \ \ \lambda_1<\lambda_2
\ee
Because of the definition of the quantum binomial coefficient,
$E^>_{\lambda_1,\lambda_2} = 0$ for $\lambda_1<\lambda_2$,
while  $E^<_{\lambda_1,\lambda_2} = 0$ for $\lambda_1\geq \lambda_2$.
This allows each of them to remain a Cherednik eigenfunction in the foreign domain
with unspecified eigenvalue. In turn, this means one can remove the requirements $\lambda_1\lessgtr \lambda_2$, and just work with two functions $E^\lessgtr_{\lambda_1,\lambda_2}$.

At $q=1$, these expressions become
\be\label{E21q1}
E^> = x_1^{\lambda_2}x_2^{\lambda_1}\left(1+\frac{x_1}{x_2}\right)^{\lambda_1-\lambda_2} =
(x_1x_2)^{\lambda_2}(x_1+x_2)^{\lambda_1-\lambda_2}
\ee
\be\label{E12q1}
E^< =x_1^{\lambda_1}x_2^{\lambda_2}\left(1+\frac{x_1}{x_2}\right)^{\lambda_2-\lambda_1-1}
= x_1^{\lambda_1}x_2^{\lambda_1+1}(x_1+x_2)^{\lambda_2-\lambda_1-1}
\ee
In this case, however, the Cherednik equations trivialize: both quantities are symmetric,
so $T_i$'s do not act on them, while the dilatation operator is unity for $q=1$.
In any case, they both have the same dimension, hence satisfy the common equation
$\left(x_1\frac{\p}{\p x_1}+x_2\frac{\p}{\p x_2}\right)F = (\lambda_1+\lambda_2)F$.

In general, the relation between the two solutions at $N=2$ is
\be
E^<_{\lambda,\lambda+a}(qx_2,x_1) = q^\lambda x_1 E^>_{\lambda+a-1,\lambda}(x_1,x_2)
= (x_1x_2)^\lambda x_1\,{\rm Pol}_{a-1}(x_1,x_2)
\ee
i.e. one is obtained from another by permutation of variables, the shift of $a$ and rescaling.
The answer is proportional to $\lambda$-independent polynomial of degree $a-1$.

The situation resembles solving the equation $\left(x\frac{\p}{\p x}\right)^2 F = \lambda^2 F$:
there are two solutions $x^\lambda$ and $x^{-\lambda}$.
In order to construct a polynomial solution at integer $\lambda$, one picks up
$F = x^{|\lambda|} = \theta(\lambda)x^\lambda + \theta(-\lambda)x^{-\lambda}$,
however, this hardly has a reasonable analytical continuation to arbitrary $\lambda$.
A solution analytic in $\lambda$ is
\be
F = C_+ \frac{x^{\lambda}}{\Gamma(\lambda)} + C_-\frac{x^{-\lambda}}{\Gamma(-\lambda)}
\ee
it still contains two free parameters.

One can rewrite formulas (\ref{E21i}), (\ref{E12i}) in the form suitable for extending them to an arbitrary pair of complex parameters $(\lambda_1,\lambda_2)$: if $\lambda_1\ge\lambda_2$ ($\lambda:=\lambda_1-\lambda_2$),
\be\label{E12}
E^>_{[\lambda_1,\lambda_2]}&=&x_1^{\lambda_2}x_2^{\lambda_1}\sum_{k=0}\left({x_1\over x_2}\right)^k q^{\lambda-k}\ \prod_{j=1}^k{(1-q^{\lambda-j+1})\over (1-q^j)}\prod_{j=0}^k{(1-tq^{j})\over(1- tq^{\lambda-j})}=\nn\\
&=&x_1^{\lambda_2}x_2^{\lambda_1}\sum_{k=0}\left({x_1\over x_2}\right)^k q^{\lambda-k}\ {(q^{\lambda-k+1};q)_k\over(q^{\lambda-k}t;q)_{k+1}}
{(t;q)_{k+1}\over(q;q)_k}=\sum_\mu x_2^{\lambda_1+\lambda_2-\mu}E^>_{[\lambda_1,\lambda_2]/[\mu]}E_\mu(x_1)
\ee
and, if $\lambda_2\ge\lambda_1$ ($\lambda:=\lambda_2-\lambda_1$),
\be\label{E21}
E^<_{[\lambda_1,\lambda_2]}&=&x_1^{\lambda_1}x_2^{\lambda_2}\sum_{k=0}\left({x_1\over x_2}\right)^k
\ \prod_{j=1}^k{(1-q^{\lambda-j})\over (1-q^j)}{(1-tq^{j-1})\over (1-tq^{\lambda-j})}=\nn\\
&=&x_1^{\lambda_1}x_2^{\lambda_2}\sum_{k=0}\left({x_1\over x_2}\right)^k{(q^{\lambda-k};q)_k\over(q^{\lambda-k}t;q)_{k}}
{(t;q)_{k}\over(q;q)_k}=\sum_\mu x_2^{\lambda_1+\lambda_2-\mu}E^<_{[\lambda_1,\lambda_2]/[\mu]}E_\mu(x_1)
\ee
since
\be
E_{[\mu]}(x)=x^\mu
\ee

This means that there are two different skew polynomials
\be\label{E2}
\begin{array}{rcll}
E^>_{[\lambda_1,\lambda_2]/[\mu]}&=&q^{\lambda_1-\mu}
\displaystyle{(q^{\lambda_1-\mu+1};q)_{\mu-\lambda_2}\over(q^{\lambda_1-\mu}t;q)_{\mu-\lambda_2+1}}
\displaystyle{(t;q)_{\mu-\lambda_2+1}\over(q;q)_{\mu-\lambda_2}}\ \ \ \ \ \ \ \ \ \ & \lambda_2\le\mu\le\lambda_1\cr
\cr
E^<_{[\lambda_1,\lambda_2]/[\mu]}&=&\displaystyle{(q^{\lambda_2-\mu};q)_{\mu-\lambda_1}\over(q^{\lambda_2-\mu}t;q)_{\mu-\lambda_1}}
\displaystyle{(qt;q)_{\mu-\lambda_1}\over(q;q)_{\mu-\lambda_1}}& \lambda_1\le\mu\le\lambda_2
\end{array}
\ee
related by the permutation of $\lambda_1$ and $\lambda_2$:
\be\label{per1}
E^>_{[\lambda_1,\lambda_2]/[\mu]}=q^{\lambda_1-\mu}{(1-q^{\lambda_1-\lambda_2})\over(1-tq^{\lambda_1-\lambda_2})}
{(1-tq^{\mu-\lambda_2})\over(1-q^{\lambda_1-\mu})}E^<_{[\lambda_2,\lambda_1]/[\mu]}
\ee
These two polynomials are immediately extended to arbitrary complex values of $\lambda_1$ and $\lambda_2$, see sec.\ref{nst}.

\subsection{Skew non-symmetric Macdonald polynomials at $N=3$}

Now we consider $N=3$ case, which is already a very typical example. The first few recursion formulas look like
\be
E_{[0,0,1]}&=&x_3 E_{[0,0]}\nn\\
E_{[0,1,0]}&=&{qt(1-t)\over (1-qt^2)}x_3 E_{[0,0]}+E_{[0,1]} \nn \\
&& \!\!\!\!\!\!\!\!\!\!\!\!\!\!\!\!\!\!\text{in more detail}, \ \
E_{[0,1,0]} =  \overbrace{\frac{( x_2 - t^{-1}x_3)}{ (x_2 - x_3)}}^{c^2_{3,2}=(23)}
\cdot \underbrace{\Xi_{010}}_{x_2\Omega}
+ \overbrace{\frac{(1 - t^{-1})(qt^2x_3 - x_2)}{(qt^2 - 1)(x_3 - x_2)}}^{c_{3,2}^3=\boxed{32}_2}
\cdot \underbrace{\Xi_{001}}_{x_3\Omega}
= {qt(1-t)\over (1-qt^2)}\cdot x_3\cdot \underbrace{ \Omega}_{E_{[0,0]}} + \underbrace{x_2\Omega}_{E_{[0,1]}}   \nn\\
E_{[1,0,0]}&=&{q(1-t)\over (1-qt)}x_3E_{[0,0]}+E_{[1,0]}\nn\\
E_{[0,1,1]}&=&x_3E_{[0,1]}\nn\\
E_{[1,0,1]}&=&x_3E_{[1,0]}-{q(1-t)^2\over(1-qt)(1-qt^2)}x_3E_{[0,1]}\nn\\
E_{[1,1,0]}&=&{q(1-t)\over(1-qt)}x_3E_{[1,0]}+{q(1-t)(1-q)\over(1-qt)^2}x_3E_{[0,1]}+E_{[1,1]}\nn\\
E_{[0,0,2]}&=&x_3^2+{(1-t)\over(1-qt)}x_3E_{[1,0]}+{(1-t)(1-q)\over(1-qt)^2}x_3E_{[0,1]}\nn\\
E_{[0,2,0]}&=&{q^2t(1-t)\over(1-q^2t^2)}x_3^2+{q^2t(1-t)^2\over(1-qt)(1-q^2t^2)}x_3E_{[1,0]}
+{q(1-t)(1-q^2t)(1-qt^2)\over(1-qt)^2(1-q^2t^2)}x_3E_{[0,1]}+E_{[0,2]}\nn\\
E_{[2,0,0]}&=&{q^2(1-t)\over(1-q^2t)}x_3^2+{q(1-q^2)(1-t)\over(1-q)(1-q^2t)}x_3E_{[1,0]}+E_{[2,0]}\nn\\
E_{[1,1,1]}&=&x_3E_{[1,1]}\nn\\
E_{[0,1,2]}&=&x_3^2E_{[0,1]}+{t(1-t)\over(1-qt^2)}x_3E_{[1,1]}\nn\\
E_{[1,0,2]}&=&x_3^2E_{[1,0]}+{(1-t)\over(1-qt)}x_3E_{[1,1]}\nn\\
E_{[0,2,1]}&=&{q(1-t)\over(1-qt)}x_3^2E_{[0,1]}+x_3E_{[0,2]}\nn\\
E_{[2,0,1]}&=&{q(1-t)\over(1-qt)}x_3^2E_{[1,0]}+{q^2(1-t)^2\over(1-qt)(1-q^2t^2)}x_3^2E_{[0,1]}+x_3E_{[2,0]}
-{q^2(1-t)^2\over(1-q^2t)(1-q^2t^2)}x_3E_{[0,2]}\nn\\
E_{[1,2,0]}&=&{q^2t(1-t)\over(1-q^2t^2)}x_3^2E_{[1,0]}+{qt(1-t)(1-q^2t)\over(1-qt)(1-q^2t^2)}x_3E_{[1,1]}+
{q(1-t)\over(1-qt)}x_3E_{[0,2]}+E_{[1,2]}\nn\\
E_{[2,1,0]}&=&{q^2(1-t)^2\over(1-qt)^2}x_3^2E_{[1,0]}+{q^2(1-t)(1-q)(1-qt^2)\over(1-qt)^2(1-q^2t^2)}x_3^2E_{[0,1]}+
{q(1-t)(1-q^2t)(1-qt^2)\over(1-qt)^2(1-q^2t^2)}x_3E_{[1,1]}+\nn\\
&+&{q(1-t)\over(1-qt)}x_3E_{[2,0]}+
{q^2(1-t)^2(1-q)\over(1-qt)^2(1-q^2t)}x_3E_{[0,2]}+E_{[2,1]}\nn\\
E_{[0,0,3]}&=&x_3^3E_{[0,0]}+{(1-t)(1-q^2)\over(1-q)(1-q^2t)}x_3^2E_{[1,0]}+{(1-t)(1-q^2)\over(1-qt)(1-q^2t)}x_3^2E_{[0,1]}+\nn\\
&+&{(1-t)\over(1-q^2t)}x_3E_{[2,0]}+{(1-t)(1-q^2)\over(1-q^2t)^2}x_3E_{[0,2]}\nn\\
E_{[0,3,0]}&=&{q^3t(1-t)\over(1-t^2q^3)}x_3^3E_{[0,0]}+{q^3t(1-t)^2(1-q^2)\over(1-q^2t)(1-q^3t^2)(1-q)}x_3^2E_{[1,0]}+
{q^2(1-t)(1-qt^2)(1-q^3t)\over(1-qt)(1-q^2t)(1-q^3t^2)}x_3^2E_{[0,1]}+\nn\\
&+&{q^3t(1-t)^2\over(1-q^2t)(1-q^3t^2)}x_3E_{[2,0]}+
{q(1-t)(1-q^2t^2)(1-q^3t)(1-q^2)\over(1-q)(1-q^2t)^2(1-q^3t^2)}x_3E_{[0,2]}+E_{[0,3]}\nn\\
E_{[3,0,0]}&=&{q^3(1-t)\over(1-q^3t)}x_3^3E_{[0,0]}+{q^2(1-t)(1-q^3)(1-qt)\over(1-q)(1-q^2t)(1-q^3t)}x_3^2E_{[1,0]}+
{q(1-t)(1-q^3)\over(1-q)(1-q^3t)}x_3E_{[2,0]}+E_{[3,0]}
\ee

The general rules one can observe are:

\begin{itemize}
\item $E_{[\lambda_1,\lambda_2,\lambda_3]/[\lambda,\mu]}=0$ if $[\lambda,\mu]$ is not within permutations of $[\lambda_1,\lambda_2,\lambda_3]$.
\item $E_{[\lambda_1,\lambda_2,\lambda_3]/[\lambda_1,\lambda_2]}=x_3^{\lambda_3}$.
\item $E_{[\lambda_1,0,\lambda_3]/[\lambda_1,\lambda_3]}=E_{[0,\lambda_2,\lambda_3]/[\lambda_2,\lambda_3]}=0$.
\item $E_{[\lambda_1,\lambda_2,\lambda_3]/[\lambda,\mu]}=E_{[\lambda_1-x,\lambda_2-x,\lambda_3-x]/[\lambda-x,\mu-x]}$ for any integer $x$(assuming that all entries in the weak compositions are non-negative).
\end{itemize}
Hence, one can put one of the entries in the weak composition zero. Then:
\be\label{3to2}
[n,k,0]=\left\{
\begin{array}{rl}
\sum_{j=k}^n\left([j,k]+\sum_{p=0}^{k-1}\Big([j,p]+[p,j]\Big)\right)&\hbox{at }n\ge k\cr\cr
\sum_{p=0}^{n}\left(\sum_{j=n}^{k-1}[j,p]+\sum_{j=n+1}^{k}[p,j]\right)&\hbox{at }n< k
\end{array}\right.
\ee
The general rule for the expansion of $E_\lambda(x_1,x_2,\ldots,x_N)$ into $x_N^{|\lambda|-|\mu|}E_\mu(x_1,x_2,\ldots,x_{N-1})$, where the length of $\lambda$ is $N$, and of $\mu$ is $N-1$, is as follows. Denote with $\lambda^{(1)}$ the largest of $\{\lambda_i\}$, $\lambda^{(2)}$ the second large, etc. Then, admissible values of $\mu_i$'s are
$\lambda^{(1)}\ge\mu_1\ge\lambda^{(2)}\ge\mu_2\ge\ldots\ge\mu_{N-1}\ge\lambda^{(N)}$ plus all possible permutations of such $\mu_i$'s. In fact, some of the coefficients of the expansion vanish when these inequalities become equalities making the inequalities strict. This is seen in the example (\ref{3to2}): in accordance with the rule, one would write $\sum_{p=0}^{k}\Big([j,p]+[p,j]\Big)$ at $n\ge k$, however, the coefficients in front of $[k,j]$ at $j\ge k$ vanish, see (\ref{E<}).

Coming to the coefficients of these expansions, consider first the Young diagram case, i.e. $n\ge k$:
\be
E_{[n,k,0]/[j,0]}=q^{n-j+k}{(1-t)\over (1-q^kt)}\prod_{i=0}^{j-k}{(1-q^it)\over(1-q^{n-i-k}t)}
\prod_{i=1}^{j-k}{(1-q^{n-i-k+1})\over(1-q^i)}
\ee
At $j\ge 1$:
\be
E_{[n,k,0]/[j,1]}=q^{n-j+k-1}{(1-q^jt^2)(1-q^nt)(1-q^k)\over(1-q^jt)(1-q^nt^2)(1-q)}
\prod_{i=1}^{k-1}{(1-q^{i-1}t)\over(1-q^{i+1}t)}\prod_{i=0}^{j-k}{(1-q^it)\over(1-q^{n-i-k}t)}
\prod_{i=1}^{j-k}{(1-q^{n-i-k+1})\over(1-q^i)}
\ee
and at $j\ge m$:
\be
E_{[n,k,0]/[j,m]}&=&q^{n-j+k-m}\prod_{i=1}^m{(1-q^{k-i+1})(1-q^it)\over(1-q^i)(1-q^{k-i}t)}
\prod_{i=0}^{n-j-1}{(1-q^{n-m-i}t^2)(1-q^{n-i}t)\over(1-q^{n-i}t^2)(1-q^{n-m-i}t)}\times\nn\\
&\times&\prod_{i=0}^{j-k}{(1-q^it)\over(1-q^{n-i-k}t)}
\prod_{i=1}^{j-k}{(1-q^{n-i-k+1})\over(1-q^i)}
\ee
Similarly, at $j<m$:
\be\label{E<}
E_{[n,k,0]/[j,m]}&=&q^{n-j}{(1-q^{j+1})\over(1-q^{n-j}t)}
\prod_{i=0}^j{(1-q^{k-i})(1-q^it)(1-q^{n-i}t)(1-q^{j-i}t^2)\over(1-q^{i+1})(1-q^{k-i}t)(1-q^{m-i}t)(1-q^{n-i}t^2)}
\times\nn\\
&\times&
\prod_{i=1}^{m-k}{(1-q^{n-i-k+1})(1-q^{i-1}t)\over(1-q^i)(1-q^{n-i-k+1}t)}
\ee
Thus, at $n\ge k\ge p$, one finally obtains
\be
E_{[n,k,p]}=\sum_{m=p}^k\sum_{j=k}^nx_3^{n+k+p-j-m}\left(E_{[n-p,k-p,0]/[j-p,m-p]}E^>_{[j,m]}(x_1,x_2)+
E_{[n-p,k-p,0]/[m-p,j-p]}E^<_{[m,j]}(x_1,x_2)\right)
\ee

In terms of the Pochhammer symbols, one can rewrite these formulas in the form that is conveniently comparable with the standard symmetric Macdonald case from \cite{Macbook,NS}:
\be\label{E32}
\hspace{-1cm}E_{[n,k,0]/[j,m]}=\left\{
\begin{array}{rl}
q^{n-j+k-m}\displaystyle{(q^{k-m+1};q)_m(q^{n-j+1};q)_{j-k-1}(q^{j+1}t;q)_{n-j}(t;q)_{m+1}(t;q)_{j-k+1}(q^{j-m+1}t^2;q)_{n-j}\over
(q;q)_m(q;q)_{j-k-1}(q^{j-m+1}t;q)_{n-j}(q^{k-m}t;q)_{m+1}(q^{n-j}t;q)_{j-k+1}(q^{j+1}t^2;q)_{n-j}}
&\hbox{at }j\ge m\cr\cr
q^{n-j}\displaystyle{(q^{k-j};q)_{j+1}(q^{n-m+1};q)_{m-k}(q^{n-j+1}t;q)_{j}(t;q)_{j+1}(t;q)_{m-k}(q^{m-j}t^2;q)_{j+1}\over
(q;q)_{j}(q;q)_{m-k}(q^{m-j}t;q)_{j+1}(q^{k-j}t;q)_{j+1}(q^{n-m+1}t;q)_{m-k}(q^{n-j}t^2;q)_{j+1}}
&\hbox{at }j< m
\end{array}
\right.\nn\\
\ee
These expressions are to be compared with (\ref{Mnn-1})
\be
M_{[n,k,0]/[j,m,0]}={(q^{j-k+1};q)_{n-j}(q^{m+1};q)_{k-m}(q^{n+1}t;q)_{n-j}(t;q)_{n-j}(t;q)_{k-m}(q^{j-m}t^2;q)_{n-j}\over
(q;q)_{n-j}(q;q)_{k-m}(q^{j-k}t;q)_{n-j}(q^mt;q)_{k-m}(q^{j-m+1};q)_{n-j}(q^nt^2;q)_{n-j}}
\ee
Here $n\ge k$, $j\ge m$.

\section{Skew non-symmetric Macdonald polynomials: generic case}

Let us rewrite formulas (\ref{E32}) in a more compact form (and with non-zero $\lambda_3$):
\be
E_{\lambda/\mu}&=&q^{|\lambda|-|\mu|}{(q^{\lambda_2-\mu_2+1};q)_{\mu_2-\lambda_3}\over(q^{\lambda_2-\mu_2}t;q)_{\mu_2-\lambda_3+1}}
{(q^{\lambda_1-\mu_1+1};q)_{\mu_1-\lambda_2}\over(q^{\lambda_1-\mu_1}t;q)_{\mu_1-\lambda_2+1}}
{(q^{\mu_1-\lambda_3+1}t;q)_{\lambda_1-\mu_1}\over(q^{\mu_1-\lambda_3+1}t^2;q)_{\lambda_1-\mu_1}}\times\nn\\
&\times&{(t;q)_{\mu_2-\lambda_3+1}\over(q;q)_{\mu_2-\lambda_3}}{(t;q)_{\mu_1-\lambda_2+1}\over(q;q)_{\mu_1-\lambda_2}}
{(q^{\mu_1-\mu_2+1}t^2;q)_{\lambda_1-\mu_1}\over(q^{\mu_1-\mu_2+1}t;q)_{\lambda_1-\mu_1}}
\ee
Here both $\lambda=[\lambda_1,\lambda_2,\lambda_3]$ and $\mu=[\mu_1,\mu_2]$ are Young diagrams. This formula has to be compared with (\ref{Mnn-1}).

This formula straightforwardly generalizes to a general formula for $E_{\lambda/\mu}$ at arbitrary $n$, with $\lambda$ and $\mu$ being Young diagrams:
\be\label{rcl}\boxed{
\begin{array}{rcl}
E_{\lambda/\mu}=E(\lambda,\mu)&:=&q^{|\lambda|-|\mu|}\prod_{i=1}^{N-1}\displaystyle{(q^{\lambda_i-\mu_i+1};q)_{\mu_i-\lambda_{i+1}}\over
(q^{\lambda_i-\mu_i}t;q)_{\mu_i-\lambda_{i+1}+1}}\times \prod_{i=1}^{N-1}\displaystyle{(t;q)_{\mu_i-\lambda_{i+1}+1}\over
(q;q)_{\mu_i-\lambda_{i+1}}}\times\cr\cr
&\times& \prod_{i<j\le N-1}\displaystyle{(q^{\mu_i-\mu_j+1}t^{j-i+1};q)_{\lambda_i-\mu_i}\over
(q^{\mu_i-\mu_j+1}t^{j-i};q)_{\lambda_i-\mu_i}}\times\prod_{i+1<j\le N}
\displaystyle{(q^{\mu_i-\lambda_j+1}t^{j-i-1};q)_{\lambda_i-\mu_i}\over(q^{\mu_i-\lambda_j+1}t^{j-i};q)_{\lambda_i-\mu_i}}
\end{array}
}
\ee
This formula also perfectly matches $E_\lambda$ in (\ref{E12}).

Note that, at $N=3$, if $\mu$ is {\bf not} a Young diagram, i.e. $\mu_1<\mu_2$, one gets
\be
E_{\lambda/\mu}=\sigma_1^{(\mu)}\left[q^{\mu_1-\lambda_2}{(1-q^{\lambda_1-\mu_1}t)(1-q^{\lambda_2-\mu_2})(1-q^{\mu_1-\mu_2}t^2)
\over(1-q^{\mu_1-\lambda_2}t)(1-q^{\mu_1-\mu_2}t)(1-q^{\lambda_1-\mu_2}t^2)}
E(\lambda,\mu) \right]
\ee
where $\sigma_i$ permutes $\mu_1$ and $\mu_2$.

In fact, this is the general claim: suppose $\lambda$ and $\mu$ are Young diagrams. Then,
\be\label{sigmamu}
E_{\lambda/\sigma_i\mu}=\sigma_i^{(\mu)}\left[q^{\mu_i-\lambda_{i+1}}{(1-q^{\lambda_i-\mu_i}t)(1-q^{\lambda_{i+1}-\mu_{i+1}})(1-q^{\mu_i-\mu_{i+1}}t^2)
\over(1-q^{\mu_i-\lambda_{i+1}}t)(1-q^{\mu_i-\mu_{i+1}}t)(1-q^{\lambda_i-\mu_{i+1}}t^2)}
E(\lambda,\mu) \right]
\ee
However, this does not work when one makes already two successive permutations.

Similarly, if $\lambda$ is {\bf not} a Young diagram: $\lambda_3\le\lambda_1<\lambda_2$, one gets
\be\label{per2}
E_{\lambda/\mu}=\sigma_1^{(\lambda)}\left[tq^{\mu_1-\lambda_2}{(1-q^{\lambda_2-\mu_2}t)(1-q^{\lambda_1-\mu_1})(1-q^{\lambda_1-\lambda_2}t)
\over(1-q^{\lambda_1-\lambda_2})(1-q^{\mu_1-\lambda_2}t)(1-q^{\lambda_1-\mu_2}t^2)}
E(\lambda,\mu) \right]
\ee
(see also (\ref{per1})), and if $\lambda_2<\lambda_3\le\lambda_1$,
\be\label{per3}
E_{\lambda/\mu}=\sigma_2^{(\lambda)}\left[q^{\mu_2-\lambda_2}{(1-q^{\lambda_2-\mu_2})(1-q^{\lambda_2-\lambda_3}t)
\over(1-q^{\lambda_2-\lambda_3})(1-q^{\mu_2-\lambda_3}t)}
E(\lambda,\mu) \right]
\ee

\section{Non-symmetric triad\label{nst}}

Note that (\ref{E12}) and (\ref{E21}) at generic complex values of $\lambda_1$, $\lambda_2$ are infinite power series. They become polynomials either at non-negative integer values of $\lambda_{1,2}$ ($\lambda_1\ge \lambda_2$ and the sum runs to $\lambda_1-\lambda_2$ in (\ref{E12}); $\lambda_2> \lambda_1$ and the sum runs to $\lambda_2-\lambda_1-1$ in (\ref{E21})), or at integer values of $m:=-\log_q t$ (the sum runs to $m-1$ in (\ref{E12}) and $m>0$, and runs to $m$ in (\ref{E21}) and $m\ge 0$).

Similarly, at $N=3$, there are 6 different branches, all of them being related by simple relations like (\ref{per1}), (\ref{per2}), (\ref{per3}). The branch obtained by continuation of (\ref{rcl}) to arbitrary complex $\lambda_i$'s in this case is given by
\be\label{NS3}
E(x_1,x_2,x_3;\lambda_1,\lambda_2,\lambda_3)=\sum_{\mu_1\ge\mu_2}x_3^{|\lambda|-|\mu|} E_{\lambda/\mu}E^>_{[\mu_1,\mu_2]}(x_1,x_2)
+\sum_{\mu_1<\mu_2} x_3^{|\lambda|-|\mu|}E_{\lambda/\sigma_1\mu}E^<_{[\mu_1,\mu_2]}(x_1,x_2)
\ee
and one has to use formulas (\ref{rcl}), (\ref{sigmamu}), (\ref{E12}), (\ref{E21}). The sums in this formula at arbitrary complex $\lambda$ run to infinity so that this $E(x_1,x_2,x_3;\lambda_1,\lambda_2,\lambda_3)$ is a power series of ratios of $x_3/x_1$, $x_3/x_2$ and $x_1/x_2$. One can similarly consider power series of differently chosen ratios of $x_1$, $x_2$ and $x_3$, and there are totally 6 possible series: each series is associated with the concrete Weyl chamber. In terms of recursion by branching rules described in sec.\ref{3.2}, formula (\ref{NS3}) is obtained by recursion 
\be 
E(x_1,x_2,x_3)=\sum E^{(3)}(1)\to E(x_1,x_2),\ \ \ \ \ \ E(x_1,x_2)\to \sum E^{(2)}(1)E(x_1)
\ee
There are six possibilities to construct similar recursion choosing different sequences of $x_i$, each giving rise to sums containing distinct ratios. This different recursions (or sets of ratios) are associated with six different Weyl chambers.

The sum of all these six series provide us with eigenfunctions of the Cherednik Hamiltonians with arbitrary eigenvalues
\be\label{ev}
\Lambda^{(i)}_{\lambda}=q^{\lambda_i}t^{N-i}
\ee
See a detailed description in \cite{NSM7}.

This consideration is immediately extended to an arbitrary $N$, and allows one to introduce a notion of {\bf non-symmetric triad}, analogous to the standard basic triad in Macdonald theory \cite{MMP3}, the latter has two polynomial reductions (to the symmetric Macdonald polynomials when $\{\lambda_i\}$'s form a partition, and to the Baker-Akhiezer function when $t=q^{-m}$, $m\in\mathbb{Z}_{\ge 0}$), the first of them gives rise to symmetric polynomials. The basic difference is that the new triad does not have a polynomial reduction to symmetric polynomials at all, hence the name non-symmetric. Instead, the non-symmetric triad (at a concrete Weyl chamber) consists of $N!$ branches, each of them having one polynomial reduction to non-symmetric Macdonald polynomials (at properly ordered non-negative integer $\{\lambda_i\}$'s) and another one to counterparts of the Baker-Akhiezer function (at integer $t=q^{-m}$ with a proper restriction). Properties of this latter are still to be studied. The sum over all Weyl chambers for any given branch gives rise to an eigenfunction of the Cherednik Hamiltonians with arbitrary eigenvalues (universal solution).

In order to illustrate how the triad looks like, we briefly describe it in the $N=2$ case. In this case, there are two branches $E^>$ and $E^<$, and one can rewrite formulas (\ref{E12}), (\ref{E21}) of sec.\ref{secN2} in the form
\be\label{NSE}
\mathfrak{E}^>(x_1,x_2;\lambda_1,\lambda_2)&\sim&x_1^{\lambda_2}x_2^{\lambda_1}\cdot t^{z_1-z_2\over 2}\cdot\sum_{k=0}\left({x_1\over x_2}\right)^k q^{\lambda-k}\ \prod_{j=1}^k{(1-tq^{j})\over (1-q^j)}{(1-t^{-1}q^{\lambda-j+1})\over (1-q^{\lambda-j})},\ \ \ \ \ \ \ \ \lambda:=\lambda_1-\lambda_2\nn\\
\mathfrak{E}^<(x_1,x_2;\lambda_1,\lambda_2)&\sim&x_1^{\lambda_1}x_2^{\lambda_2}\cdot t^{z_1-z_2\over 2}\cdot\sum_{k=0}\left({x_1\over x_2}\right)^k \
\prod_{j=1}^k{(1-tq^{j-1})\over (1-q^j)}{(1-t^{-1}q^{\lambda-j})\over (1-q^{\lambda-j})},\ \ \ \ \ \ \ \lambda:=\lambda_2-\lambda_1
\ee
where we made the shifts $q^{\lambda_i}\to q^{\lambda_i}t^{i-{3\over 2}}$ in the first expression and $q^{\lambda_i}\to q^{\lambda_i}t^{{3\over 2}-i}$ in the second one, and use the notation $x_i:=q^{z_i}$. We assume here $\lambda_{1,2}$ are arbitrary complex parameters so that the sums run unbounded.

To compare, the standard Noumi-Shiraishi function at $N=2$ (in the other Weyl chamber as compared with \cite{NS,MMP3}) is
\be
\NS_{q,t}(x_1,x_2;\lambda_1,\lambda_2)=x_2^{\lambda_1}x_1^{\lambda_2}\cdot t^{z_1-z_2\over 2}\cdot
\sum_{k=0}\left({x_1\over x_2}\right)^k\prod_{j=1}^{k}{(1-tq^{j-1})
\over (1-q^j)}{(1-t^{-1}q^{\lambda-j+1})
\over (1-q^{\lambda-j})},\ \ \ \ \ \ \ \lambda:=\lambda_1-\lambda_2
\ee

The both power series (\ref{NSE}) are {\bf not} related by the formula that is an immediate corollary of (\ref{B})\footnote{Formula (\ref{B}) is guaranteed by the identity
$$
\prod_{j=1}^{\lambda-k}{(1-q^{\lambda-j+1})\over (1-q^j)}\prod_{j=0}^{\lambda-k}{(1-tq^{j})\over(1- tq^{\lambda-j})}=
\prod_{j=1}^k{(1-q^{\lambda-j+1})\over (1-q^j)}{(1-tq^{j-1})\over (1-tq^{\lambda-j+1})}
$$
which is applicable to (\ref{E12}), (\ref{E21}) at natural values of $\lambda$.
}
:
\be
\mathfrak{E}^<(x_1,x_2;y_1,y_2)\sim x_2\mathfrak{E}^>(qx_2,x_1;(qt)^{-1}y_2,t^{-1}y_1)
\ee
where we introduced the variables $y_i:=q^{\lambda_i}$. This is since $\mathfrak{E}^<(x_1,x_2;y_1,y_2)$ is a power series in ${x_1\over x_2}$, and $\mathfrak{E}^>(x_2,x_1;y_2,y_1)$ is a power series in ${x_2\over x_1}$. Moreover, the both power series (\ref{NSE}) {\bf are not} eigenfunctions of the Cherednik Hamiltonians. This is again because these are the series in powers of ${x_1\over x_2}$, and the Cherednik Hamiltonians relate expansions in ${x_1\over x_2}$ and in ${x_2\over x_1}$ since they contain the permutation operator $\sigma_1$.

In order to construct eigenfunctions with arbitrary eigenvalues, one has to use the combinations \cite{NSM7}:
\be\label{65}
{\cal E}^>(x_1,x_2;\lambda_1,\lambda_2)=\mathfrak{E}^>_m\Big(x_1,x_2;\lambda_1,\lambda_2\Big)+{q^{\lambda_1+1}\over t^{1\over 2}x_1}\mathfrak{E}^<_m\Big(x_2,x_1/q;\lambda_2,\lambda_1+1\Big)\nn\\
{\cal E}^<(x_1,x_2;\lambda_1,\lambda_2)=\mathfrak{E}^<_m\Big(x_1,x_2;\lambda_1,\lambda_2\Big)+t^{-{1\over 2}}q^{1-\lambda_2}x_2\mathfrak{E}^>_m\Big(qx_2,x_1;\lambda_2-1,\lambda_1\Big)
\ee
which {\bf are} eigenfunctions of the Cherednik Hamiltonians with the same eigenvalues
\be
\Lambda^{(i)}(\lambda_1,\lambda_2)=q^{\lambda_i}t^{1\over 2}
\ee
This allows one to interpret them as two branches of the same eigenfunction or as a two-component vector of solutions to the eigenfunction equation. They also satisfy
\be
{\cal E}^<(x_1,x_2;y_1,y_2)\sim x_2{\cal E}^>(qx_2,x_1;(qt)^{-1}y_2,t^{-1}y_1)
\ee
The second terms in (\ref{65}) are just, correspondingly, series $\mathfrak{E}^>_m$ and $\mathfrak{E}^<_m$ in the toher Weyl chamber.

Now there are two polynomial reductions of (\ref{NSE}) \cite{NSM7}: if one puts for the first and second lines in (\ref{NSE}) accordingly $q^{\lambda_i}=q^{\mu_i}t^{\rho_i}$ and $q^{\lambda_i}=q^{\mu_i}t^{-\rho_i}$ ($\rho_i={N+1\over 2}-i$ are the components of the Weyl vector) with $\{\mu_i\}$ being non-negative integers, and if one puts $t=q^{-m}$, $m\in\mathbb{Z}_{\ge 0}$. In the first case, one obtains the corresponding non-symmetric Macdonald polynomials, and, in the second case, (a counterpart of) the Baker-Akhiezer function, which is a (quasi)polynomial at arbitrary $\lambda_i$'s being a finite sum of $m+1$ terms. Further, putting $t=q^{-m}$ in the respective non-symmetric Macdonald polynomial, one sees that it has two groups of terms: the first group is equal to one of the BA function, while the second one, to the other BA function (with permuted arguments), which is expressed in (\ref{65}). In the symmetric case, this second group was obtained from the first one by the Weyl group action.

\section{Conclusion}

In the paper, we considered branching rules for the non-symmetric Macdonald polynomials which allow to construct these polynomials recursively in the number of particles $N$. The non-symmetric Macdonald polynomials are eigenfunctions of the Cherednik Hamiltonians with eigenvalues parameterized by the weak compositions. It turns out that the resulting answers obtained for these polynomials with this procedure can be immediately extended to formal power series, sums of which over all Weyl chambers are eigenfunctions of the Cherednik Hamiltonians with arbitrary complex eigenvalues. This is much similar to the Noumi-Shiraishi power series \cite{NS}, which provides eigenfunctions of the Ruijsenaars-Schneider Hamiltonians with arbitrary complex eigenvalues.

This allowed us to construct a non-symmetric triad, which is a counterpart of the basic triad \cite{MMP3}. The triad has two polynomial reductions: to non-symmetric Macdonald polynomials (at peculiar eigenvalues) and to multivariable Baker-Akhiezer function (at peculiar $t=q^{-m}$, $m\in\mathbb{Z}_{\ge 0}$).

In fact, in variance with the Noumi-Shiraishi case, at each Weyl chamber, there are $N!$ different eigenfunction power series and triads, all of them being related by formulas like (\ref{per1}), (\ref{per2}), (\ref{per3}). At the peculiar eigenvalues corresponding to the non-symmetric Macdonald polynomials, these $N!$ triads are associated with $N!$ permutations of lines of the Young diagram parameterizing the polynomials (and expressions in distinct Weyl chambers coincide). In fact, these triads can be treated as different branches of one large triad.

In this paper, as a reference example, we write down an explicit formula for the coefficients of the triad branch associated with the Young diagram: formula (\ref{rcl}). We postpone a full list of formulas as well as a detailed discussion of the structure of triad branches to the forthcoming publications. We also discuss elsewhere properties of the Baker-Akhiezer function emerging upon the reduction of the triad at $t=q^{-m}$. At last, we notice that actually our consideration can be further promoted to eigenfunctions of the twisted Cherednik operators, as well as to eigenfunctions of DAHA Hamiltonians associated with other root systems (symmetric and non-symmetric Koornwinder polynomials) \cite{BMP}, but these stories were also well beyond the scope of the paper, and will be presented elsewhere.

\section*{Acknowledgements}

This work is supported by the RSF grant 26-12-00191.

\newpage

\section{Appendix: Some explicit expressions}

Here we sketch some explicit formulas, concerning non-symmetric polynomials $E_\lambda$ for low level $L:=\sum_{i=1}^N\lambda_i$ and $N$.
They can turn useful for visualizing the emerging structures and for their future analysis.

\bigskip

{\bf Level [1]:}
\be
E^{L=1}_{N,1}&=&
E_{0,\ldots,0,1} =\underbrace{\Xi_{0,\ldots,0,0,1}}_{\Xi_{N,1}=\tilde\Xi^{(N)}_N},
\nn \\ \nn \\ \nn \\
E^{L=1}_{N,2}&=&
E_{0,\ldots,0,1,0} = \frac{x_{N-1}-t^{-1}x_N }{x_{N-1}-x_N}\cdot
\underbrace{\Xi_{0,\ldots,0,1,0}}_{\Xi_{N,2}=\tilde \Xi^{(N)}_{N-1}}
+ \frac{(1-t^{-1})}{(qt^{N-1}-1)}\cdot \frac{qt^{N-1}x_N - x_{N-1}}{x_N-x_{N-1}}
\cdot \underbrace{\Xi_{0,\ldots,0,0,1}}_{\Xi_{N,1}=\tilde \Xi^{(N)}_N},
\nn \\ \nn \\ \nn \\
E^{L=1}_{N,3}&=&
E_{0,\ldots,0,1,0,0} = \frac{(x_{N-2}-t^{-1}x_N)(x_{N-2}-t^{-1}x_{N-1}) }{(x_{N-2}-x_N)(x_{N-2}-x_{N-1}}\cdot
\underbrace{\Xi_{0,\ldots,0,1,0,0}}_{\Xi_{N,3}=\tilde \Xi^{(N)}_{N-2}} + \nn \\
&+& \frac{1-t^{-1}}{qt^{N-2}-1}\left( \frac{(x_{N-1}-t^{-1}x_N)\cdot(qt^{N-2}x_{N-1}-x_{N-2})}{(x_{N-1}-x_N)(x_{N-1}-x_{N-2})}\cdot
\underbrace{\Xi_{0,\ldots,0,1,0}}_{\Xi_{N,2}=\tilde \Xi^{(N)}_{N-1}}
+
\frac{(x_N-t^{-1}x_{N-1})\cdot (qt^{N-2}x_N - x_{N-2})}{(x_N-x_{N-1})(x_N-x_{N-2})}
\cdot \underbrace{\Xi_{0,\ldots,0,0,1}}_{\Xi_{N,1}=\tilde \Xi^{(N)}_N}\right),
\nn \\ \nn \\ \nn \\
E^{L=1}_{N,4}&=&
E_{0,\ldots,0,1,0,0,0} = \frac{(x_{N-3}-t^{-1}x_N)(x_{N-3}-t^{-1}x_{N-1})(x_{N-3}-t^{-1}x_{N-2}) }{(x_{N-3}-x_N)(x_{N-3}-x_{N-1})(x_{N-3}-x_{N-2})}\cdot
\underbrace{\Xi_{0,\ldots,0,1,0,0,0}}_{\Xi_{N,4}} + \nn \\
&+& \frac{1-t^{-1}}{qt^{N-3}-1}\left(
 \frac{(x_{N-2}-t^{-1}x_N)(x_{N-2}-t^{-1}x_{N-1})\cdot(qt^{N-3}x_{N-2}-x_{N-3}) }{(x_{N-2}-x_N)(x_{N-2}-x_{N-1})(x_{N-2}-x_{N-3})}\cdot
\underbrace{\Xi_{0,\ldots,0,1,0,0}}_{\Xi_{N,3}}
+ \right. \nn \\ &+&\left.
\frac{(x_{N-1}-t^{-1}x_N)(x_{N-1}-t^{-1}x_{N-2})\cdot(qt^{N-3}x_{N-1}-x_{N-3})}
{(x_{N-1}-x_N)(x_{N-1}-x_{N-2})((x_{N-1}-x_{N-3})}\cdot
\underbrace{\Xi_{0,\ldots,0,1,0}}_{\Xi_{N,2}}
+ \right. \nn \\ &+& \left.
\frac{(x_N-t^{-1}x_{N-1})(x_N-t^{-1}x_{N-2})\cdot(qt^{N-3}x_N - x_{N-3})}
{(x_N-x_{N-1})(x_N-x_{N-2})(x_N-x_{N-3})}
\cdot \underbrace{\Xi_{0,\ldots,0,0,1}}_{\Xi_{N,1}}\right),
\nn
\ee
\be
\ldots
\ee
In general,
\be
 E^{L=1}_{N,k} := E_{0 \ldots 0\underbrace{10 \ldots 0}_k} = \ \
\Xi_{N,k} \cdot \prod_{j=0}^{k-2} \frac{x_{N-k+1}-t^{-1}x_{N-j}}{x_{N-k+1}-x_{N-j}}
+ \nn \\
+  \frac{1-t^{-1}}{qt^{N-k+1}-1}  \cdot\sum_{i=0}^{k-2} \Xi_{N,i+1} \cdot
\frac{qt^{N-k+1}x_{N-i}-x_{N-k+1}}{x_{N-i}-x_{N-k+1}}
 \prod_{j\neq i}^{k-2}\frac{x_{N-i}-t^{-1}x_{N-j}}{x_{N-i}-x_{N-j}}
\ee
or
\be
\boxed{
\tilde E^{L=1}_{N,k} := E_{\underbrace{0 \ldots 01}_k 0 \ldots 0} = \ \
\tilde\Xi^{[1]}_{k} \cdot \prod_{j=k+1}^{N} \frac{x_{k}-t^{-1}x_{j}}{x_{k}-x_{j}}
+  \frac{1-t^{-1}}{qt^{k}-1}  \cdot\sum_{i=k+1}^{N} \tilde \Xi^{[1]}_i \cdot
\frac{qt^{k}x_{i}-x_{k}}{x_{i}-x_{k}}
 \prod_{k<j\neq i}^{N}\frac{x_{i}-t^{-1}x_{j}}{x_{i}-x_{j}}
}
\ee
From now on, we suppress $N$ in the notation.

Convenient building blocks are
\be
\eta_{\rho,i,j}:=\frac{\rho x_i-x_j}{x_i-x_j} = \frac{1-\rho\frac{x_i}{x_j}}{1-\frac{x_i}{x_j}},\nn \\
\zeta_{\rho,i,k}:=\prod_{\stackrel{j=k}{j\neq i}}^N \frac{\eta_{\rho,i,j}}{\rho}
\ee
or
\be
(ij) := \frac{tx_i-x_j}{t(x_i-x_j)}= \frac{\eta_{ij}(t)}{t} = \eta_{ji}(t^{-1}), \nn \\
\boxed{ij}_k:= \frac{1-t^{-1}}{qt^k-1}\cdot \frac{qt^kx_i-x_j}{x-I-x_j}
= \frac{1-t^{-1}}{qt^k-1}\cdot\eta_{ij}(qt^k)
\ee

Then
\be
\boxed{
\tilde E^{L=1}_{N,k} := E_{\underbrace{0 \ldots 01}_k 0 \ldots 0} = \ \
\tilde\Xi^{[1]}_{k} \cdot \prod_{j=k+1}^{N} \frac{\eta_{k,j}(t)}{t}
+  \frac{1-t^{-1}}{qt^{k}-1}  \cdot\sum_{i=k+1}^{N} \tilde \Xi^{[1]}_i \cdot
\eta_{ i,k}(qt^k)\cdot \zeta_{t,i,k+1}(t)
} \nn \\
= \sum_{i\geq k}^N c_{N,k}^i \Xi^{[1]}_i
= \Xi^{[1]}_k \prod_{j>k}^N (kj) + \sum_{i>k}^N \Xi^{[1]}_i\cdot  \boxed{ik}_k \prod_{\stackrel{j>k}{j\neq i}}^N (ij)
\ee
Each term in the sum contains equal number $(N-k)$ of $\eta$-factors.

For future comparison with other levels we convert this formula into a sequence of tables:

$N=2$:
\be
c_{2,k}^i =  \
\begin{array}{|c||c|c|} \hline   \phantom._{k}\backslash \phantom.^{i}
& \Xi_{1} & \Xi_{2} \\  \hline\hline && \\
E_1 & (12) & \cre{\boxed{21}_1} \\ && \\ \hline && \\
E_2 & & 1 \\ && \\ \hline
\end{array}
\ee

$N=3$:
\be
c_{3,k}^i = \
\begin{array}{|c||c|c|c|} \hline \phantom._{k}\backslash \phantom.^{i}
& \Xi_{1} & \Xi_{2} & \Xi_{3} \\  \hline\hline &&& \\
E_{1}=& (12)(13)  &(23)\cre{\boxed{21}_1}  & \cre{\boxed{31}_1(32)}  \\ &&& \\
E_{2}= && (23) & \cre{\boxed{32}_2}   \\ &&&   \\
E_{3}= &&& 1 \\ &&& \\ \hline
\end{array}
\ee

$N=4$:
\be
c_{4,k}^i = \
\begin{array}{|c||c|c|c|c| } \hline  \phantom._{k}\backslash \phantom.^{i}
& \Xi_{1} & \Xi_{2} & \Xi_{3} & \Xi_{4}   \\ \hline\hline &&&&  \\
E_{1}=& (12)(13)(14)   &(23)(24)\cre{\boxed{21}_1}    &  (34)\cre{\boxed{31}_1(32)}
& \cre{\boxed{41}_1(42)(43)}     \\  &&&& \\
E_{2}= && (23)(24) & (34)\cre{\boxed{32}_2}  & \cre{\boxed{42}_2(43)}   \\ &&&&  \\
E_{3} = &&& (34) & \cre{\boxed{43}_3}     \\ &&&& \\
E_{4}= &&&& 1    \\ &&&&\\ \hline
 \end{array}
\ee

$N=5$:
\be
c_{5,k}^i = \
\begin{array}{|c||c|c|c|c|c| } \hline \phantom._{k}\backslash \phantom.^{i}
& \Xi_{1} & \Xi_{2} & \Xi_{3} & \Xi_{4} & \Xi_5   \\ \hline\hline &&&&&  \\
E_{1}=& (12)(13)(14)(15)   &(23)(24)(25)\cre{\boxed{21}_1}    &  (34)(35)\cre{\boxed{31}_1(32)}
& (45)\cre{\boxed{41}_1(42)(43)} &  \cre{\boxed{51}_1(52)(53)(54)}   \\  &&&&& \\
E_{2}= && (23)(24)(25) & (34)(35)\cre{\boxed{32}_2}  & (45)\cre{\boxed{42}_2(43)} & \cre{\boxed{52}_2 (53)(54)} \\ &&&&&  \\
E_{3} = &&& (34)(35) & (45)\cre{\boxed{43}_3}  & \cre{\boxed{53}_3(54)}   \\ &&&&& \\
E_{4} = &&&& (45) & \cre{\boxed{54}_4}     \\ &&&&& \\
E_{5}= &&&&& 1    \\ &&&&&\\ \hline
 \end{array}
 \nn
\ee

Clearly, the red blocks in the tables are {\it stable}: do not change, only increase in size  with $N$.

\bigskip

{\bf Level [1,1]:}

\be
E^{[1,1]}_{k_1,k_2} :=E_{\underbrace{\underbrace{0 \ldots 01}_{k_1} 0 \ldots 0 1}_{k_2} 0 \ldots 0} = \ \
\sum_{\stackrel{ i_1<i_2}{i_1\geq k_1, i_2\geq k_2} }^N c_{N,k_1,k_2}^{i_1,i_2}\cdot\Xi^{[1,1]}_{i_1,i_2}
\ee

$N=2$:   $$E_{12}=\Xi_{12}$$

$N=3$:
\be
\begin{array}{c||c|c|c|} \phantom._{k_1,k_2}\backslash \phantom.^{i_1,i_2}
& \Xi_{12} & \Xi_{13} & \Xi_{23} \\  \hline\hline &&& \\
E_{12}=& (13)(23)  &(12)\cre{\boxed{32}_1}  & \cre{\boxed{31}_1(21)}  \\ &&& \\
E_{13}= && (12) & \cre{\boxed{21}_2}   \\ &&& \\ \hline &&& \\
E_{23}= &&& 1 \\ &&& \\ \hline
\end{array}
\ee

$N=4$:
\be
\begin{array}{c||c|c|c|c|c|c|}  \phantom._{k_1,k_2}\backslash \phantom.^{i_1,i_2}
& \Xi_{12} & \Xi_{13} & \Xi_{14} & \Xi_{23} & \Xi_{24} & \Xi_{34} \\ \hline\hline &&&&&& \\
E_{12}=& (13)(14)(23)(24)  &(12)(14)(34)\cre{\boxed{32}_1}  & (12)(13)\cb{\boxed{42}_1(43)}
& (24)(34)\cre{\boxed{31}_1(21)} &(23)\cb{\boxed{41}_1(21)(43)} & \cb{c_{4,12}^{34}}  \\  &&&&&&\\
E_{13}= && (12)(14)(34) & (12)(13)\cb{\boxed{43}_2}  & (24)(34)\cre{\boxed{21}_2}
& (23)\cb{\boxed{21}_2\boxed{43}_2} & \cb{c_{4,13}^{34}} \\ &&&&&& \\
E_{14} = &&& (12)(13) & \cb{0} & (23)\cb{\boxed{21}_2}  & \cb{\boxed{31}_2(32)}  \\ &&&&&&\\  \hline &&&&&&\\
E_{23}= &&&& (24)(34) & (23)\cb{\boxed{43}_2}  & \cb{\boxed{42}_2 (32)} \\ &&&&&&\\
E_{24}= &&&&& (23) & \cb{\boxed{32}_3} \\ &&&&&&\\ \hline &&&&&&\\
E_{34}= &&&&&& 1 \\ &&&&&&\\ \hline
\end{array}
\nn
\ee

\be
c_{4,12}^{34} = \frac{(q-1)(qt^2-1)}{t(qt-1)^2}\cdot\boxed{41}_1\boxed{32}_2+\boxed{31}_1\boxed{42}_1(32)(41)
= \boxed{31}_1\boxed{42}_2(21)(32) + \boxed{32}_1\boxed{41}_2(12)(31), \nn \\
c_{4,13}^{34} = \boxed{41}_1(31)(42) -\boxed{23}_1\boxed{31}_2\boxed{42}_2
\label{c34}
\ee

\bigskip

$N=5:$
{\footnotesize
\be
\begin{array}{c||c|c|c|c|c}
& \Xi_{12} & \Xi_{13} & \Xi_{14} &  \Xi_{15} & \ldots
\\ \hline\hline &&&&& \\
E_{12}=& (13)(14)(15)(23)(24)(25)  &(12)(14)(15)(34)(35)\cre{\boxed{32}_1}
&(12)(13)(15)(45)\cb{\boxed{42}_1 (43)} & (12)(13)(14)\co{\boxed{52}_1 (53)(54)}   \\ &&&&&  \\
E_{13}=&& (12)(14)(15)(34)(35)  &(12)(13)(15)(45)\cb{\boxed{43}_2} &(12)(13)(14)\co{\boxed{53}_2 (54)}  \\ &&&&&  \\
E_{14}=&&& (12)(13)(15)(45)  &(12)(13)(14)\co{\boxed{54}_3}   \\ &&&&&  \\
E_{15}=&&&& (12)(13)(14)   &  \\ &&&&&  \\ \hline   &&&&&  \\
%
\end{array}
\nn
\ee

\bigskip

\be
\!\!\!\!\!\!\!\!\!\!\!\!\!\!\!\!\!\!\!\!\!\!\!\!\!\!\!\!\!\!\!\!\!\!\!\!\!\!
\begin{array}{c||c|c|c|c|c|c|}
& \Xi_{23} & \Xi_{24} & \Xi_{25} & \Xi_{34} & \Xi_{35} & \Xi_{45} \\ \hline\hline &&&&&& \\
E_{12}=&(24)(25)(34)(35)\cre{\boxed{31}_1(21)}
&(23)(25)(45)\cb{\boxed{41}_1(21) (43)}&(23)(24)\co{\boxed{51}_1(21)(54)(53)}
&(35)(45)\cdot \cb{c_{4,12}^{34}} &(34)\cdot\co{\tilde c_{5,12}^{35} \cdot(54)}
&\co{\tilde c_{5,12}^{45}\cdot(43)(53)}\\ &&&&&&  \\
E_{13}=& (24)(25)(34)(35)\cre{\boxed{21}_2}&(23)(25)(45)\cb{\boxed{21}_2\boxed{43}_2}
&(23)(24)\co{\boxed{21}_2\boxed{53}_2(54)}
&(35)(45)\cdot \cb{c_{4,13}^{34}} &(34)\cdot \co{\tilde c_{5,13}^{35}\cdot(54)}&\co{c_{5,13}^{45}} \\ &&&&&&  \\
E_{14}=& \cb{0} &(23)(25)(45)\cb{\boxed{21}_2} &(23)(24)\co{\boxed{21}_2\boxed{54}_3}
&(35)(45)\cb{\boxed{31}_2(32)}&(34)\co{\boxed{31}_2\boxed{54}_3(32)}&\co{c_{5,14}^{45}}  \\ &&&&&&  \\
E_{15}=& \co{0} &\co{0} &(23)(24)\co{\boxed{21}_2}&\co{0}&(34)\co{\boxed{31}_2(32)}&\co{\boxed{41}_2 (42)(43)}
\\ &&&&&&  \\ \hline &&&&&&  \\
E_{23}=&  (24)(25)(34)(35) &(23)(25)(45)\cb{\boxed{43}_2}&(23)(24)\co{\boxed{53}_2(54)}
&(35)(45)\cb{\boxed{42}_2(32)}&(34)\co{\boxed{52}_2(54)(32)} &\co{c_{5,23}^{45}} \\ &&&&&&  \\
E_{24}=& &  (23)(25)(45) &(23)(24)\co{\boxed{54}_3}&(35)(45)\cb{\boxed{32}_3}
&(34)\co{\boxed{32}_3\boxed{54}_3} &\co{c_{5,24}^{45}}   \\ &&&&&&  \\
E_{25}=&&&  (23)(24)  &\co{0}&(34)\co{\boxed{32}_3} & \co{\boxed{42}_3(43)} \\ &&&&&&  \\ \hline   &&&&&&  \\
E_{34}=&&&&  (35)(45)   & (34)\co{\boxed{54}_3} & \co{\boxed{53}_3(43)}  \\ &&&&&&  \\
E_{35}=&&&&&  (34)   & \co{\boxed{43}_4}  \\ &&&&&&  \\ \hline   &&&&&& \\
E_{45}=&&&&&&  1   \\ &&&&&& \\ \hline
\end{array}
\nn
\ee
}

\be
\tilde c_{5,12}^{35}
= \frac{(q-1)(qt^2-1)}{t(qt-1)^2}\boxed{31}_2\boxed{52}_2+\boxed{32}_1\boxed{51}_1(31)(52)
= \boxed{31}_1\boxed{52}_2(21)(32) + \boxed{32}_1\boxed{51}_2(12)(31), \nn \\
\tilde c_{5,13}^{35} = \boxed{51}_1(31)(52)-\boxed{23}_1\boxed{31}_2\boxed{52}_2,
\nn \\ \nn \\
\tilde c_{5,12}^{45} = \boxed{41}_1\boxed{52}_2(21)(42) + \boxed{42}_1\boxed{51}_2(12)(41),  \nn \\
c_{5,13}^{45} = \boxed{51}_2\boxed{43}_2(53)(52)(45)+\boxed{53}_2\boxed{41}_2(54)(43)(42), \nn \\
c_{5,14}^{45} = \boxed{51}_2(45)(52)(53)+\boxed{41}_2\boxed{54}_3(42)(43),  \nn \\
c_{5,23}^{45} = \frac{(qt-1)(qt^3-1)}{t(qt^2-1)^2}\boxed{43}_3\boxed{52}_3+\boxed{42}_2\boxed{53}_2(43)(52)
=   \boxed{43}_2\boxed{52}_3(42)(23)+\boxed{42}_2\boxed{53}_3(43)(32), \nn \\
c_{5,24}^{45} = \boxed{52}_2(42)(53) - \boxed{34}_2\boxed{42}_3\boxed{53}_3
\label{c54}
\ee

\bigskip

Properties:
\begin{itemize}
\item{}
The number of $\eta$-factors in $E_{k_1,k_2}$ is $(N-k_1-1)+(N-k_2)=2N-k_1-k_2-1$.

\item{}
Elements $c_{N,k_1,k_2}^{i_1,i_2}$ are non-vanishing only for $i_1<i_2$ and $i_1 \geq k_1$ and  $i_2\geq k_2$.
In particular, they vanish for $k_1<i_1<i_2<k_2$, therefore
$c_{4,1,4}^{2,3}= c_{5,1,4}^{2,3}=c_{5,1,5}^{2,3}=c_{5,1,5}^{2,4}=c_{5,1,5}^{3,4}=c_{5,2,5}^{3,4}=0$
in the tables.

\item{}
Diagonal element $c_{N,i_1,i_2}^{i_1,i_2}$ is a product
\be
c_{N,k_1,k_2}^{k_1,k_2}=
c^0_{N,k_1,k_2}:=\frac{1}{(k_1,k_2)}\prod_{i_1>k_1}\prod_{i_2>k_2} (k_1i_1)(k_2 i_2)
\ee
and it persists as a factor in all entries above,  $c_{N,k_1,k_2}^{i_1,i_2}$ with $k_1\leq i_1$ and $k_2\leq i_2$.
We put them in front, so that explanation/rule  is needed only for the squares and brackets to the right of them,
which form the ratio $\frac{c_{N,k_1,k_2}^{i_1,i_2}}{c_{N,i_1,i_2}^{i_1,i_2}}$.
These ratios possess a  stability property:
\be
\boxed{
\frac{c_{N,k_1,k_2}^{i_1,i_2}}{c_{N,i_1,i_2}^{i_1,i_2}}  \ \
\text{ does not depend on} \  N
}
\ee
even if this ratio is not factorized.
This is easy to see, comparing the factorized quantities in different tables.
We mark them in colors: red for those appearing at $N=3$ and further,
blue for  $N\geq 4$, orange for $N\geq 5$ etc.
Less trivial examples are provided by $c_{5,12}^{34} = (35)(45)\cdot c_{4,12}^{34}$ and
$c_{5,12}^{34} = (35)(45)\cdot c_{4,12}^{34}$ in the last  table,
where $c_{4,12}^{34}$ and $c_{4,13}^{34}$ are already non-trivial expressions.
At level one these invariant blocks were all marked by red.

\item{}
Non-vanishing elements of the first sub-diagonal with $i_1+i_2=k_1+k_2+1$ are
\be
c_{N,k_1,k_2}^{k_1,k_2+1}= \left\{\begin{array}{ccc}
c^0_{N,k_1,k_2+1} \cdot \boxed{k_2+1,k_2}_{k_2-1} & {\rm for} & (k_1,k_2)\neq (N-2,N) \\ &&\\
c_{N,N-2,N}^{N-1,N} = \boxed{N-1,N-2}_{N-1}
\end{array}\right.
\ee

\item{}
There are sum rules:

Since for $a=0$  the polynomials $\Xi$ are just $\Xi_\alpha = \prod_i q^{\frac{\alpha_i(\alpha_i-1)}{2}}  x_i$
\be
\sum_{i_1<i_2} c_{N,k_1,k_2}^{i_1,i_2} x_{i_1}x_{i_2} = E^{(0)}_{k_1,k_2}
\label{sumpol}
\ee
for all $k_1<k_2$ are polynomials in $x$, known as non-symmetric Macdonald polynomials.

\item{}
In addition,
\be
\sum_{i=k}^N c_{N,k}^{[1],i}  = \frac{1}{t^{N-k}}\cdot \frac{qt^N-1}{qt^k-1}, \nn \\
\sum_{i_1<i_2}^N c_{N,k_1,k_2}^{[1,1],i_1,i_2} =
\frac{1}{t^{2N-1-k_1-k_2}} \frac{(qt^N-1)(qt^{N-1}-1)}{(qt^{k_1+1}-1)(qt^{k_2-1}-1)}, \nn \\
\sum_{i=k}^N c_{N,k}^{[2],i}  = \frac{1}{qt^{2N-1-k}}\cdot \frac{(qt^N-1)(q^2t^N-1)}{(qt-1)(q^2t^k-1)}, \nn \\
\sum_{i=k}^N c_{N,k}^{[3],i}  = \frac{1}{q^3t^{3N-2-k}}\cdot
\frac{(qt^N-1)(q^2t^N-1)(q^3t^N-1)}{(qt-1)(q^2t-1)(q^3t^k-1)}, \nn \\
\ldots , \nn \\
\sum_{i=k}^N c_{N,k}^{[r],i}  = \frac{1}{q^rt^{rN+1-r-k}}\cdot
\frac{(q^r t^N-1)}{(q^rt^k-1)}\prod_{j=1}^{r-1} \frac{(q^jt^N-1) }{(q^jt-1)}, \nn \\
\ldots
\label{sumrule}
\ee
i.e. the sums of coefficients $c$ {\it per se}, i.e. with $\Xi_\alpha=1$ are constants (independent of $x$).
Formally this corresponds to putting $a=\infty$.

Both these sum rules are non-trivial, because $c_{N,k_1,k_2}^{i_1,i_2}$ are rational expressions.

\item{}
(\ref{sumrule}) provides a decomposition of the simplest non-factorized coefficients
(like $c_{4,12}^{3,4}, c_{4,13}^{34}, c_{5,14}^{45}, c_{5,23}^{45}, c_{5,24}^{45}$)
into the sums of factorized ones, but with a constant free term from the r.h.s. of (\ref{sumrule}).
 Expressions like
(\ref{c34}) and (\ref{c54})  are much better.

\end{itemize}

{\bf Level 2:}

\be
\tilde E^{L=2}_{N} =  E_{0 \ldots 0 2} = \ \
\frac{1}{q} \tilde\Xi^{[2]}_{N} \prod_{j=1}^{N-1} \frac{x_N-(qt)^{-1}x_j}{x_N-q^{-1}x_j}
+ \frac{t-1}{qt-1}\sum_{i=1}^{N-1} \tilde\Xi^{[1,1]}_{i,N}\cdot \frac{x_i -t^{-1}x_N }{x_i-qx_N }
\prod_{j\neq i}^{N-1} \frac{x_i-t^{-1}x_j}{x_i-x_j}
\ee
Here we can use another set of building blocks:
\be
u_i(a,b):=\prod_{1\leq j\neq i}^N \frac{x_i-a^{-1}x_j}{x_i-b^{-1}x_j} \\
v_{k,i}:=\left\{\begin{array}{ccc} 1 & {\rm if} & k=i \\
\frac{1-t^{-1}}{q^2t^k-1}\cdot \frac{q^2t^kx_i-x_k}{x_i-x_k} &{\rm if} & k\neq i
\end{array}\right. \\
w_{k,i}:=\prod_{k+1=j\neq i} \frac{x[i]-t^{-1}x_j}{x_i-x_j}
\ee

Our previous results for the sector $[1,1]$, which is closed by itself,
can now be embedded into more general tables for level 2.
For this purpose we introduce a new notation
\be
[i,j]_{a,b}:=q^at^bx_i-x_j
\ee
Our old $(i,j) = \frac{1}{t} \frac{[i,j]_{0,1}}{[i,j]_{0,0}}$
and $\boxed{i,j}_b = \frac{(t-1)}{t(qt^b-1)}\frac{[i,j]_{1,b}}{[i,j]_{0,0}}$.

Then

\bigskip

$N=2$:
\be
\begin{array}{c||c|c|c|} \phantom._E \backslash \phantom.^\Xi
& \Xi^{[2]}_{20} & \Xi^{[2]}_{02} & \Xi^{[11]}_{12} \\  \hline\hline &&& \\
E^{[2]}_{20}=& \frac{1}{qt^2}\cdot\frac{[12]_{11}[12]_{01}}{[12]_{10}[12]_{00}}
&\frac{1}{qt}\cdot\frac{[21]_{11}}{[21]_{10}}
\cdot \cg{ \frac{1-t^{-1}}{q^2t-1}\cdot\frac{[21]_{21}} {[21]_{00}} }
& \cg{\frac{(q+1)(t-1)}{t^2(q^2t-1)}\cdot\frac{[21]_{11}[12]_{01}}{[21]_{10}[12]_{10}}}  \\
&&& \\
E^{[2]}_{02}= && \frac{1}{qt}\cdot\frac{[21]_{11}}{[21]_{10}}
&  \cg{-\frac{1-t^{-1}}{qt-1}\cdot\frac{[12]_{01}}{[21]_{10}}}   \\ &&& \\
\hline &&& \\
E^{[11]}_{11}= &&& 1 \\ &&& \\ \hline
\end{array}
\ee

\bigskip

Green color marks ingredients which first appear at $N=2$ and then persist at all other $N$.
Red marks the items, appearing for the first time at $N=3$.

\bigskip

$N=3$:

\be
\begin{array}{c||c|c|c|c} \phantom._E \backslash \phantom.^\Xi
& \Xi^{[2]}_{1} & \Xi^{[2]}_{2} & \Xi^{[2]}_{3} &
\\  \hline\hline &&&& \\
E^{[2]}_{200}=& \frac{1}{qt^4}\cdot\frac{[12]_{11}[13]_{11}[12]_{01}[13]_{01}}{[12]_{10}[13]_{10}[12]_{00}[13]_{00}}
& \frac{1}{qt^3}
\cdot \frac{[21]_{21} [23]_{11}[23]_{01}} {[21]_{10}[23]_{10} [23]_{00}}\cdot\cg{\frac{1-t^{-1}}{q^2t-1}\frac{[21]_{11}}{[21]_{00}} }
&  \frac{1}{qt^2}\cdot\frac{[32]_{11}[31]_{11}}{[32]_{10}[31]_{10}}
\cdot\cre{\frac{1-t^{-1}}{q^2t-1}\cdot\frac{[31]_{21}[32]_{01}}{[31]_{00}[32]_{00}}}
& \\
&&&& \\
E^{[2]}_{020}= && \frac{1}{qt^3}
\cdot \frac{[21]_{11} [23]_{11}[23]_{01}} {[21]_{10}[23]_{10} [23]_{00}}
&  \frac{1}{qt^2}\cdot\frac{[32]_{11}[31]_{11}}{[32]_{10}[31]_{10}}
\cdot\cre{\frac{1-t^{-1}}{q^2t-1}\cdot\frac{[32]_{22}}{[32]_{00}}}
& \\
&&&& \\
E^{[2]}_{002}= &&& \frac{1}{qt^2}\cdot\frac{[32]_{11}[31]_{11}}{[32]_{10}[31]_{10}}
& \\ &&&& \\
\hline &&&&
\end{array}
\nn
\ee

\be
\!\!\!\!\!\!\!\!\!\!\!\!\!\!\!\!\!\!
\begin{array}{c||c|c|c||c| } \phantom._E \backslash \phantom.^\Xi
& \Xi^{[11]}_{12} & \Xi^{[11]}_{13} & \Xi^{[11]}_{23}&{\rm sum\ rule}
\\  \hline\hline &&&& \\
E^{[2]}_{200}=
& (13)(23)\cdot\cg{\frac{(q+1)(t-1)}{t^2(q^2t-1)}\frac{[21]_{11}[12]_{01}}{[21]_{10}[12]_{10}}}
& (12) \cdot \cre{\frac{(q+1)(t-1)}{t^3(q^2t-1)}\frac{[31]_{11}[13]_{01}[32]_{01}}{[31]_{10}[13]_{10}[32]_{00}} }
& \cre{\frac{(q+1)(t-1)^2}{t^4(q^2t-1)(qt-1)}
\frac{[21]_{11}[31]_{11}[32]_{01}[23]_{01}}{[21]_{00}[31]_{00}[32]_{10}[23]_{10}}  }
&\frac{(q^2t^3-1)(qt^3-1)}{qt^4(q^2t-1)(qt-1)} \\
&&&& \\
E^{[2]}_{020}=
& (13)(23)\cdot\cg{\frac{-(1-t^{-1})}{qt-1}\frac{[12]_{01}}{[21]_{10}}}
& (12)\cdot \cre{\frac{(t-1)^2}{t^2(q^2t^2-1)(qt-1)}\frac{[32]_{22}[13]_{01}}{[31]_{10}[23]_{00}}}
& \cre{\frac{(t-1)}{t^3(q^2t^2-1)(qt-1)}\frac{[23]_{01}
\tilde c_{3,020}^{011}}
{[23]_{10}[32]_{10}[12]_{00}[31]_{00}}}
& \frac{(q^2t^3-1)(qt^3-1)}{qt^3(q^2t^2-1)(qt-1)} \\
&&&&\\
E^{[2]}_{002}=
&\cre{0} & (12)\cdot \cre{\frac{-(1-t^{-1})}{qt-1}\frac{[13]_{01}}{[31]_{10}}}
& \cre{\frac{t-1}{t^2(qt-1)} \frac{[21]_{01}[23]_{01}}{[12]_{00}[32]_{10}}}
& \frac{qt^3-1}{qt^2(qt-1)}
\\ &&&&  \\
\hline &&&& \\
E^{[11]}_{110}=& (13)(23)  &(12)\cre{\boxed{32}_1}  & \cre{\boxed{31}_1(21)}& \frac{qt^3-1}{t^2(qt-1)}  \\
&&&& \\
E^{[11]}_{101}= && (12) & \cre{\boxed{21}_2} & \frac{qt^3-1}{t(qt^2-1)}  \\ &&&& \\
E^{[11]}_{011}= &&&  1 & 1 \\ &&&& \\ \hline
\end{array}
\nn
\ee

\be
\tilde c_{3,020}^{011} = t(q+1)(qt-1)[32]_{11}[12]_{00}[13]_{00}
+ (t-1)(q+1)[31]_{22}[32]_{01}[21]_{00}
- (t-1)[31]_{22}[12]_{01}[23]_{10}
\ee

Diagonal terms are:
\be
cd_k^{[2]} = \frac{1}{qt^{2N-k-1}} \prod_{i\neq k} \frac{[ki]_{11}}{[ki]_{10}}
\prod_{i>k} \frac{[ki]_{01}}{[ki]_{00}}
\ee
and for arbitrary symmetric diagram $[r]$
\be
cd_k^{[r]} = \frac{1}{q^{\frac{r(r-1)}{2}}\cdot t^{r(N-1)k+1}}
\prod_{i<k}  \prod_{r'=1}^{r-1} \frac{[k,i]_{r',1}}{[k,i]_{r',0}}
\prod_{i>k} \prod_{r'=0}^{r-1} \frac{[r',i]_{0,1}}{[r',i]_{0,0}}
\ee

\bigskip

{\bf Comparison of different levels $r$ at $N=2$:}

Now the labeling of $E$ and $\Xi$ will be different to distinguish, say, between
$E^{[12]}_{\ldots 2\ldots 1 \ldots }$ and $E^{[12]}_{\ldots 1\ldots 2 \ldots }$.

\bigskip

{\bf Level $r=1$:}
\be
\begin{array}{c||c|c| } \phantom._E \backslash \phantom.^\Xi
& \Xi^{[1]}_{10} & \Xi^{[1]}_{01}   \\  \hline\hline &&  \\
E^{[1]}_{10}=& \frac{1}{t} \frac{[12]_{01} }{ [12]_{00}}
& \cre{\frac{1-t^{-1}}{qt-1}\frac{[21]_{11}}{[21]_{00}}}
 \\ && \\
E^{[1]}_{01}= && 1
    \\ &&  \\ \hline
\end{array}
\ee

\bigskip

{\bf Level $r=2$:}
\be
\begin{array}{c||c|c|c|} \phantom._E \backslash \phantom.^\Xi
& \Xi^{[2]}_{20} & \Xi^{[2]}_{02} & \Xi^{[11]}_{12} \\  \hline\hline &&& \\
E^{[2]}_{20}=& \frac{1}{qt^2} \frac{[12]_{11}[12]_{01}}{[12]_{10}[12]_{00}}
&\frac{1}{qt} \frac{[21]_{11}}{[21]_{10}}\cdot \cre{\frac{1-t^{-1}}{q^2t-1}\frac{[21]_{21}}{[21]_{00}}}
& \frac{q(q+1)}{qt^2}\frac{t-1}{q^2t-1} \frac{[21]_{11}[21]_{01}}{[21]_{10}[12]_{10}}  \\ &&& \\
E^{[2]}_{02}= && \frac{1}{qt} \frac{[21]_{11}}{[21]_{10}}
& \frac{q}{qt}\frac{t-1}{qt-1} \frac{[12]_{01}}{[21]_{10}}   \\ &&& \\ \hline &&& \\
E^{[11]}_{11}= &&& 1 \\ &&& \\ \hline
\end{array}
\ee

\bigskip

{\bf Level $r=3$:}

\bigskip

{\footnotesize
\be
\!\!\!\!\!\!\!\!\!\!\!\!\!\!\!\!\!\!\!\!\!\!\!\!\!\!\!\!\!
\begin{array}{c||c|c|c|c|} \phantom._E \backslash \phantom.^\Xi
& \Xi^{[3]}_{30} & \Xi^{[3]}_{03} & \Xi^{[12]}_{21}& \Xi^{[12]}_{12} \\  \hline\hline &&&& \\
E^{[3]}_{30}=& \frac{1}{q^3t^3} \frac{[12]_{21}[12]_{11}[12]_{01}}{[12]_{20}[12]_{10}[12]_{00}}
& \frac{1}{q^3t^2} \frac{[21]_{21}[21]_{11}}{[21]_{20}[21]_{10}}
\cdot \cre{\frac{1-t^{-1}}{q^3t-1}\frac{[21]_{31}}{[21]_{00}}}
&\frac{1}{qt} \frac{[12]_{01}}{[12]_{00}}\cdot
\frac{(q^2+q+1)(1-t^{-1})}{qt(q^3t-1)}\frac{[21]_{11}\cdot[12]_{11}}{[21]_{10}\cdot[12]_{20}}
&\frac{q(q^2+q+1)}{q^3t^3}\frac{(t-1)(qt-1)}{(q^3t-1)(q^2t-1)} \frac{[21]_{21}[21]_{11}\cdot[12]_{01}}
{[21]_{20}[21]_{00}\cdot[12]_{10} }  \\ &&&& \\
E^{[3]}_{03}= && \frac{1}{q^3t^2} \frac{[21]_{21}[21]_{11}}{[21]_{20}[21]_{10}}
& \frac{1}{qt} \frac{[12]_{01}}{[12]_{00}}\cdot \frac{-1}{q}\frac{1-t^{-1}}{q^2t-1}\frac{[12]_{11}}{[21]_{10}}
& \frac{q(q+1)}{q^3t^2}\frac{t-1}{q^2t-1} \frac{[21]_{11}\cdot [12]_{01}}{[21]_{20}\cdot[12]_{00}} \\ &&&& \\ \hline &&&& \\
E^{[21]}_{21}= &&&\frac{1}{qt} \frac{[12]_{01}}{[12]_{00}}
& \frac{1}{qt}\frac{t-1}{qt-1} \frac{[21]_{11}}{[21]_{00}} \\ &&&& \\
E^{[12]}_{12}= &&&& 1 \\ &&&& \\ \hline
\end{array}
\nn
\ee
}

\bigskip

\bigskip

{\bf Level $r=4$:}
\be
\begin{array}{c||c|c|c} \phantom._E \backslash \phantom.^\Xi
& \Xi^{[4]}_{40} & \Xi^{[4]}_{04}   \\  \hline\hline &&& \\
E^{[4]}_{40}=& \frac{1}{q^6t^4} \frac{[12]_{31}[12]_{21}[12]_{11}[12]_{01}}{[12]_{30}[12]_{20}[12]_{10}[12]_{00}}
&\frac{1}{q^6t^3} \frac{[21]_{31}[21]_{21}[21]_{11}}{[21]_{30}[21]_{20}[21]_{10}}
\cdot\cre{\frac{1-t^{-1}}{q^4t-1}\frac{[21]_{41}}{[21]_{00}}}
   \\ &&& \\
E^{[4]}_{04}= && \frac{1}{q^6t^3} \frac{[21]_{31}[21]_{21}[21]_{11}}{[21]_{30}[21]_{20}[21]_{10}}
   \\ &&&
\end{array}
\ee
{\footnotesize
\be
\!\!\!\!\!\!\!\!\!\!\!\!\!\!\!\!\!\!\!\!\!\!\!\!\!\!\!\!\!\!\!\!\!\!\!\!\!\!\!\!\!\!\!\!
\begin{array}{c||c|c|c|} \phantom._E \backslash \phantom.^\Xi
& \Xi^{[13]}_{31}& \Xi^{[13]}_{13}& \Xi^{[22]}_{22} \\  \hline\hline &&& \\
E^{[4]}_{40}=& \frac{1}{q^3t^2}  \frac{[12]_{11}[12]_{01}}{[12]_{10}[12]_{00}} \cdot
\frac{(q+1)(q^2+1)}{q^2t}\frac{1-t^{-1}}{q^4t-1}
\frac{[21]_{11}\cdot[12]_{21} }{[21]_{10}\cdot[12]_{30} }
&\frac{1}{q^3t}  \frac{[21]_{11}}{[21]_{10}}
\cdot\frac{(q+1)(q^2+1)}{q^2t^2}\frac{(1-t^{-1})(qt-1)}{(q^4t-1)(q^3t-1)}
 \frac{[21]_{31} [21]_{21} \cdot[12]_{01}}{[21]_{30} [21]_{00}\cdot [12]_{10}}
& \frac{q^2(q^2+q+1)(q^2+1)}{q^6t^4}\frac{t-1}{q^4t-1}
 \frac{[21]_{21}[21]_{11}\cdot[12]_{11}[12]_{01}}{[21]_{20}[21]_{10}\cdot[12]_{20}[12]_{10}}
  \\ &&& \\
E^{[4]}_{04}= & \frac{1}{q^3t^2} \frac{[12]_{11}[12]_{01}}{[12]_{10}[12]_{00}}
\cdot\frac{-1}{q^2}\frac{1-t^{-1}}{q^3t-1}\frac{[12]_{21} }{[21]_{10} }
& \frac{1}{q^3t}  \frac{[21]_{11}}{[21]_{10}}
\cdot\frac{(q^2+q+1)}{q^2t}\frac{1-t^{-1}}{q^3t-1}
 \frac{[21]_{21} \cdot[12]_{01}}{[21]_{30} \cdot[12]_{00}}
& -\frac{q(q^2+q+1)}{q^6t^3}\frac{(t-1)(qt-1)}{(q^3t-1)(q^2t-1)}
 \frac{[21]_{11}\cdot [12]_{11}[12]_{01}}{[21]_{20}[21]_{10}\cdot[12]_{10}}
\\ &&& \\ \hline &&&  \\
E^{[13]}_{31}= &\frac{1}{q^3t^2}  \frac{[12]_{11}[12]_{01}}{[12]_{10}[12]_{00}}
& \frac{1}{q^3t} \frac{[21]_{11}}{[21]_{10}}
\cdot \frac{1-t^{-1}}{q^2t-1}\cdot \frac{[21]_{21} }{ [21]_{00}}
& \frac{q(q+1)}{q^3t^2}\frac{t-1}{q^2t-1} \frac{[21]_{11}\cdot[12]_{01}}{[21]_{10}\cdot[12]_{10}}
\\ &&&  \\
E^{[13]}_{13}= &&\frac{1}{q^3t}  \frac{[21]_{11}}{[21]_{10}}
& -\frac{1}{q^3t}\frac{t-1}{qt-1} \frac{[12]_{01}}{[21]_{10}} \\ &&& \\ \hline
&&& \\
E^{[22]}_{22}= &&& 1 \\ &&& \\ \hline
\end{array}
\nn
\ee
}

\bigskip

The colored elements are
\be
\frac{c^{[r],2}_{N=2,1}}{c^{[r],2}_{N=2,2}} = \cre{\frac{1-t^{-1}}{q^rt-1}\frac{[21]_{r1}}{[21]_{00}}}
\ee


\begin{thebibliography}{12}

\bibitem{DI} J. Ding, K. Iohara, 
Lett. Math. Phys. {\bf 41} (1997) 181-193, q-alg/9608002

\bibitem{M} K. Miki, J. Math. Phys. {\bf 48} (2007) 123520

\bibitem{Che} I. Cherednik, 
IMRN (Duke M.J.) {\bf 9} (1992) 171-180

\bibitem{RS} S.N.M. Ruijsenaars, H. Schneider, Ann.Phys. (NY), {\bf 170} (1986) 370\\
S.N.M. Ruijsenaars, Comm.Math.Phys. {\bf 110} (1987) 191-213

\bibitem{Chebook} I. Cherednik, {\sl Double affine Hecke algebras},
  Vol. {\bf 319}, Cambridge University Press, 2005

\bibitem{NSM1} A.~Mironov, A.~Morozov, A.~Popolitov,
arXiv:2512.24811

\bibitem{NSM2} A.~Mironov, A.~Morozov, A.~Popolitov,
Nucl. Phys. \textbf{B1028} (2026) 117513,
arXiv:2601.10500

\bibitem{NSM3} A.~Mironov, A.~Morozov, A.~Popolitov,
Phys. Lett. \textbf{B877} (2026) 140457,
arXiv:2601.19878

\bibitem{NSM4} A.~Mironov, A.~Morozov and A.~Popolitov,
Phys. Lett. \textbf{B879} (2026) 140592,
arXiv:2602.21120

\bibitem{K} M. Kapranov,  
Algebraic geometry {\bf 7}, J.Math. Sci. {\bf 84} (1997) 1311-1360, alg-geom/9604018

\bibitem{BS} I. Burban, O. Schiffmann, 
Duke Math. J. {\bf 161} (2012) 1171, arXiv:math/0505148

\bibitem{S}  O. Schiffmann, 
J. Algebraic Combin. {\bf 35} (2012) 237-26, arXiv:1004.2575

\bibitem{Feigin} B. Feigin, M. Jimbo, T. Miwa, E. Mukhin,
Commun. Math. Phys. \textbf{356}  (2017) 285, arXiv:1603.02765

\bibitem{MMP} A. Mironov, A. Morozov, A. Popolitov, JHEP, \textbf{09} (2024) 200,
  arXiv:2406.16688

\bibitem{dFK1} P. Di Francesco, R. Kedem, Comm. Math. Phys. {\bf 369(3)} (2019) 867-928, arXiv:1704.00154

\bibitem{CF} O. Chalykh, M. Fairon, 
J.Geom.Phys. {\bf 121} (2017) 413-437,
  arXiv:1704.05814

\bibitem{MMP1} A.~Mironov, A.~Morozov, A.~Popolitov,
Phys. Lett. \textbf{B863} (2025) 139380,
arXiv:2410.10685

\bibitem{ChE}   O. Chalykh, P. Etingof, Advances in Mathematics, {\bf 238} (2013) 246-289,
arXiv:1111.0515

\bibitem{Opd95} E.M. Opdam, 
Acta Mathematica, {\bf 175(1)} (1995) 75–121

\bibitem{Mac96} I.G. Macdonald, 
Asterisque-Societe Mathematique de France, {\bf 237} (1996) 189-208

\bibitem{Che95} I. Cherednik, 
IMRN, {\bf 1995(10)} (1995) 483, q-alg/9505029

\bibitem{Cha} O. Chalykh,  Adv.Math. {\bf 166(2)} (2002) 193-259, math/0212313

\bibitem{NS} M. Noumi, J. Shiraishi, 
arXiv:1206.5364

\bibitem{MMP3} A.~Mironov, A.~Morozov, A.~Popolitov,
Phys. Lett. \textbf{B869} (2025) 139840,
arXiv:2411.16517

\bibitem{dFK2} P.~Di Francesco, R.~Kedem,
Selecta Math. \textbf{30} (2024) no.2, 23,
arXiv:2112.09798

\bibitem{MRS} J. Sekiguchi, 
Publ. RIMS, Kyoto Univ. {\bf 12} (1977) 455-459\\
A. Debiard, 
C.R. Acad. Sci. Paris (sir. I) {\bf 296} (1983) 529-532\\
S.N.M. Ruijsenaars, Comm.Math.Phys. {\bf 110} (1987) 191-213\\
I.G. Macdonald, 
Springer Lecture Notes {\bf 1271} (1987) 189-200

\bibitem{Macbook} I.G. Macdonald, {\it Symmetric functions and Hall polynomials},   Oxford University Press, 1995

\bibitem{KS} F. Knop, S. Sahi, 
Invent. Math. {\bf 128} (1997) 9-22, q-alg/9610016

\bibitem{HHL} J. Haglund, M. Haiman, N. Loehr, Am.J.Math. {\bf 130(2)} (2008) 359-383, math/0601693

\bibitem{NSM7} A. Mironov, A. Morozov, A. Popolitov, to appear

\bibitem{BMP} L. Bishler, A. Mironov, A. Popolitov, 
arXiv:2607.06738

\end{thebibliography}
\end{document}